\documentstyle[12pt,fullpage,epsfig]{article}
\def\beq{\begin{equation}}
\def\eeq{\end{equation}}

\begin{document}
\begin{center}
{\Large{\bf Studying Self-Organized Criticality with Exactly
Solved Models}}\\[2cm]

{\large{\bf Deepak Dhar}}\\
Department of Theoretical Physics, \\ 
Tata Institute of Fundamental Research, \\ 
Homi Bhabha Road, Mumbai 400~005, INDIA\\ [2cm]

\end{center} 
\bigskip 
\begin{abstract}
These lecture-notes are intended to provide a pedagogical introduction to
the abelian sandpile model of self-organized criticality, and its related
models. The abelian group structure of the algebra of particle addition
operators, the burning test for recurrent states, equivalence to the
spanning trees problem are described. The exact solution of the directed
version of the model in any dimension, and determination of the the
exponents for avalanche distribution are explained. The model's
equivalence to Scheidegger's model of river basins, Takayasu's aggregation
model and the voter model is discussed. For the undirected case, the exact
solution in 1-dimension and on the Bethe lattice is briefly described.
Known results about the two dimensional case are summarized.  
Generalization to the abelian distributed processors model is discussed,
with the Eulerian walkers model and Manna's stochastic sandpile model as
examples. I conclude by listing some still-unsolved problems.

\end{abstract}

\section{Introduction}

These notes are a somewhat expanded version of lectures given at the EPF
Lausanne under the Troisieme Cycle de la Suisse Romande programme in
November 1998, and later at the 12th Chris Engelbrecht Summer School held
at Stellenbosch in January-February 1999. The lectures were intended to
provide a pedagogical introduction to the abelian sandpile model and
related models. Here, I have tried to fill in some of the details left out
in the oral presentations. Even so, the treatment is not self- contained,
and algebraic details have sometimes been omitted in favor of citation to
original papers.  It is hoped that these notes will be useful as a
starting point for students wanting to learn about the subject in detail,
and also to others only seeking an overview of the subject.

In these notes, the main concern is the analysis of a particular model of
self-organized criticality (SOC): the so-called abelian sandpile model
(ASM), and its related models: the voter model, the $q \rightarrow 0$
limit of the $q$-state Potts model, Scheidegger's model of river networks,
the Eulerian walkers model, the abelian distributed processors model etc.
The main appeal of these models is that they are analytically tractable.
One can explicitly calculate many quantities of interest, such as
properties of the steady state, and some critical exponents, without too
much effort. As such, they are very useful for developing our
understanding of the basic principles and mechanisms underlying the
general theory. A student could think of the ASM as a `base camp' for the
explorations into the uncharted areas of non-equilibrium statistical
mechanics. The exact results in this case can also serve as proving
grounds for developing approximate treatments for more realistic problems.

While these models are rather simple to define, and in some sense soluble,
they are non-trivial, and cannot be said to be well-understood yet. For
example, it has not been possible so far to determine the critical
exponents for avalanche distributions for the undirected sandpile model in
two or three dimensions. The fairly large analytical and numerical effort
devoted to this question in two dimensions has produced only partial
results. There are many questions still unanswered. (Some will be
discussed later in the lectures.)  One may hope that these lectures will
encourage further work in this area, which will lead to a better
understanding of some of them.

Our focus here will be on the mathematical development of  these models.
However, it is useful to start with a brief discussion of their origin
as simplified models of physical phenomena.

The term SOC was first coined by Bak, Tang and Wiesenfeld (BTW) in their
well-known paper in 1987 \cite{btw87}. They  observed that many
naturally occuring phenomena display fractal behavior. Examples include
mountain ranges, river networks, coastlines, etc. The word `fractal
structure' here  means that some correlation functions show
non-trivial power law behavior. For example, in the case of mountain 
ranges, the irregular height profile can be characterized by how the
difference of height $\Delta h(R)$, between two points separated
by a distance $R$ varies with $R$. It is found that  
\begin{equation} 
 \left<[\Delta h(R)]^2\right> \sim R^x\ , 
\end{equation}
where $\left<\right>$ denotes averaging over different spatial points at
the fixed horizontal separation $R$, and $x$ is some non-trivial power.
From simple physical considerations, we see that $0\le x\le2$. When measured,
the exponent $x$ seems to vary little between one mountain range and
another.

In the case of river networks, the fractal structure can be characterized
in terms of the empirical Hack's law \cite{dodds}. This law describes
how the catchment area of any particular stream in a river basin grows as
we go down-stream along the river. If $A$ is the catchment area, and
$\ell$ is the length of the principal stream up to this point, Hack's law
states that on the average $A$ grows as $\ell^y$, where $y$ is some
exponent, $y \simeq 1.6$.

In other examples the fractal behavior is found not in the spatial
structure itself, but is manifest the power-law dependence  between some
physical observables. For example, in the case of 
earthquakes,  the frequency of earthquake of total energy $E$ is
found to vary as $E^{-z}$, where $z$ is a number close to 2, for many
decades of the energy range. (This is the well-known Gutenberg-Richter law
\cite{gutenberg}). An example in a  similar spirit is that of fluid
turbulence where the fractal behavior is observed in the way mean squared
velocity difference scales with distance, or in the spatial fractal
structure of regions of high dissipation. Many time-series like electrical
noise, stock market price variations etc.\ show power-law tails in their
power spectra [`$1/f$' noise]. For an overview of application of the SOC
idea to different natural systems, see \cite{hownatureworks}.

\begin{figure}
\begin{center}
\leavevmode
\psfig{figure=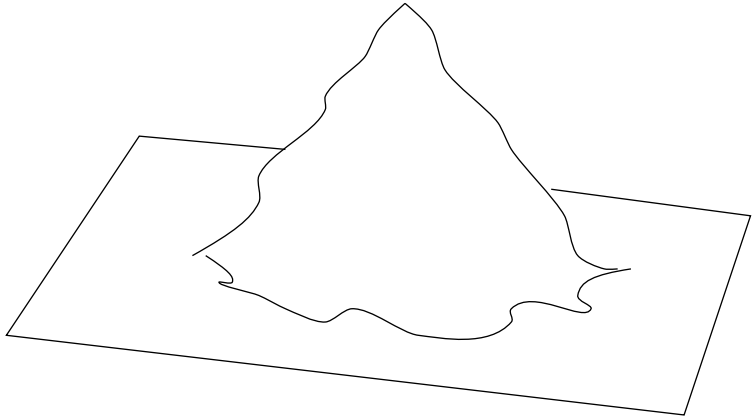,width=10cm,angle=0}
\caption{ A small sandpile on a flat table. }
\end{center}
\end{figure}

Systems exhibiting such correlations with power law decay over a wide
range of length scales are said to have critical correlations. This is
because correlations much larger than the length-scale of interactions
were first studied in equilibrium statistical mechanics in the
neighborhood of a critical phase transition. In order to observe such
critical phenomena in equilibrium systems, one needs to fine-tune some
physical parameters (such as the temperature and pressure) to specific
critical values, something rather unlikely to occur in a naturally
occuring process such as the growth of a mountain range and its erosion.
The systems we want to study may be said to be in a steady state as while
there is variation in time, average properties are roughly unchanged with
time. However, these systems are not in equilibrium: they are open and
disssipative systems which require input of energy from outside at a
constant rate to offset the dissipation. We define such states to be {\it
non-equilibrium steady states}.

BTW argued that the dynamics which give rise to the robust power-law
correlations seen in the non-equilibrium steady states in nature must not
involve any fine-tuning of parameters. It must be such that the systems
under their natural evolution are driven to a state at the boundary
between the stable and unstable states. Such a state then shows long range
spatio-temporal fluctuations similar to those in equilibrium critical
phenomena.

Bak et. al. proposed a simple example of a system whose natural dynamics
drives it towards, and then maintains it, at the edge of stability: a
sandpile. It is observed that for dry sand, one can characterize its
macroscopic behavior in terms of an angle $\theta_c$, called the angle of
repose, which depends on the detailed structure (size, shapes and
roughness etc.) of the constituting grains \cite{footnote1}. If we make a
sandpile in which the local slope is smaller than $\theta_c$ everywhere,
such a pile is stable. On such a pile, addition of a small amount of sand
will cause only a weak response, but adding a small amount of sand to a
configuration where the average slope is larger than $\theta_c$, will
often result in an avalanche whose size is of the order of the system size.
In a pile where the average slope is $\theta_c$, the response to addition
of sand is less predictable. It might cause almost no relaxation, or it
may cause avalanches of intermediate sizes, or a catastrophic avalanche
which affects the entire system. Such a state is critical. BTW
observed that if one builds a sandpile on a finite table by pouring it very
slowly, the system is invariably driven towards its critical state, and
thus organizes itself into a critical state: it shows SOC.

The steady state of this process is characterized by the following
property: Sand is being added to the system at a constant small rate, but
it leaves the system in a very irregular manner, with long periods of
apparent inactivity interspersed by events which may vary in size and
which occur at unpredictable intervals (see fig. 2). 

\begin{figure}
\begin{center}
\includegraphics[height=12cm,angle=-90] {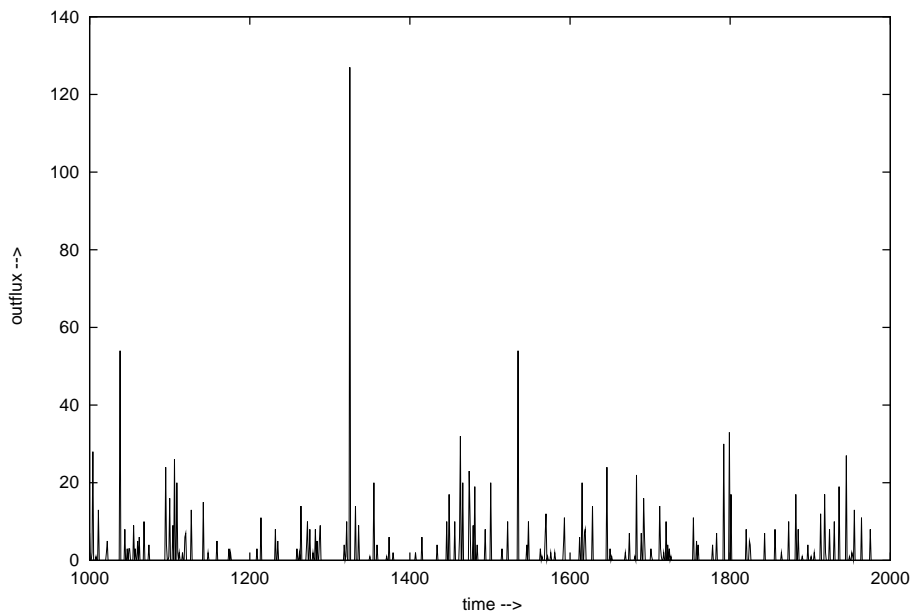}
\label{sandsch} 
\caption{ A schematic representation of the outflux of sand as a function
of time. The data shown was actually generated on computer by simulating
the BTW model of sandpiles on a square lattice of size $100 \times 100$ }
\end{center}
\end{figure}

A similar behavior is seen in earthquakes, where the build-up of stress
due to tectonic motion of the continental plates is a slow steady process,
but the release of stress occurs sporadically in bursts of various sizes.

Some people have objected to the term SOC, arguing that in fact
self-organized systems are not so different from ordinary critical
systems, because here the driving rate is the parameter which is being
fine-tuned to zero. We need not worry too much about this terminological
dispute, and prefer to adopt the point of view that fine tuning to zero is
not `unnatural', is easily realized experimentally, and often occurs in
nature. 

The first step in the study of SOC would be the precise mathematical
formulation of some simple model which exhibits it. From this introduction
it follows that such a system should have the necessary features of SOC
systems: it should be {\it extended, open, dissipative and non-linear}.
All these properties makes such systems hard to treat analytically
using the tools familiar to theoretical physicists.

\section{The BTW sandpile model}

In their original paper \cite{btw87}, Bak Tang and Wiesenfeld also
proposed a simple cellular automaton model of sandpile growth. The model
is defined on a lattice, which we take for simplicity to be the two
dimensional square lattice. There is a positive integer variable at each
site of the lattice, called the height of the sandpile at that site.  The
system evolves in discrete time.

The rules of evolution are quite simple: At each time step a site is
picked randomly, and its height $z_i$ is increased by unity. If its height
is then larger than a critical height $z_c=4$, this site is said to be
unstable. It relaxes by toppling whereby four sand grains leave the site,
and each of the four neighboring sites gets one grain. If there is any
unstable site remaining, it too is toppled. In case of toppling at a site
at the boundary of the lattice, grains falling `outside' the lattice are
removed from the system. This process continues until all sites are
stable.

Then another site is picked randomly, its height increased, and so on. To
make the rules unambiguous, the toppling of sites which were rendered
unstable during the same time step is defined to be carried out in
parallel.  It is easy to show that this process must converge to a stable
configuration in a finite number of time steps on any finite lattice using
the diffusive nature of each relaxation step (proof omitted).

The following example illustrates the toppling rules. Let the lattice size
be $4 \times 4$, and suppose at some time step the following configuration
is reached:
$$
\begin{tabular}{|c|c|c|c|} \hline
4 & 2 & 4 & 3 \\ \hline
2 & 3 & 4 & 4 \\ \hline
4 & 1 & 2 & 2 \\ \hline
3 & 1 & 3 & 4 \\ \hline
\end{tabular}
$$
We now add a grain of sand at a randomly selected site: let us say the
site on the second row from the top and the third column from the left.
Then it will reach height 5, become unstable and topple to reach
$$\mbox{ \begin{tabular}{|c|c|c|c|} \hline
4 & 2 & 4 & 3 \\ \hline
2 & 3 & 5 & 4 \\ \hline
4 & 1 & 2 & 2 \\ \hline
3 & 1 & 3 & 4 \\ \hline
\end{tabular}}
\longrightarrow
\mbox{ \begin{tabular}{|c|c|c|c|} \hline
4 & 2 & 5 & 3 \\ \hline
2 & 4 & 1 & 5 \\ \hline
4 & 1 & 3 & 2 \\ \hline
3 & 1 & 3 & 4 \\ \hline
\end{tabular}}\quad, $$
and further toppling results in 
$$
\mbox{ \begin{tabular}{|c|c|c|c|} \hline
4 & 3 & 1 & 5 \\ \hline
2 & 4 & 3 & 1 \\ \hline
4 & 1 & 3 & 3 \\ \hline
3 & 1 & 3 & 4 \\ \hline
\end{tabular}}
\longrightarrow
\mbox{ \begin{tabular}{|c|c|c|c|} \hline
4 & 3 & 2 & 1 \\ \hline
2 & 4 & 3 & 2 \\ \hline
4 & 1 & 3 & 3 \\ \hline
3 & 1 & 3 & 4 \\ \hline
\end{tabular}}\quad .
$$

In this configuration all sites are stable. One speaks in this case
of an event of size $s=4$, since there were 4 topplings. Other measures of
event size are the number of distinct sites toppled (in this case also 4),
the number of time steps needed (3 in this case), and the diameter of the
affected area (3). Suppose that one measures the relative frequency of
event sizes. A typical distribution would be characterized by a long power
law tail, with an eventual cutoff determined by the the system size $L$
(see fig. 3).

\begin{figure}
\begin{center}
\includegraphics[height=10cm,angle=-90] {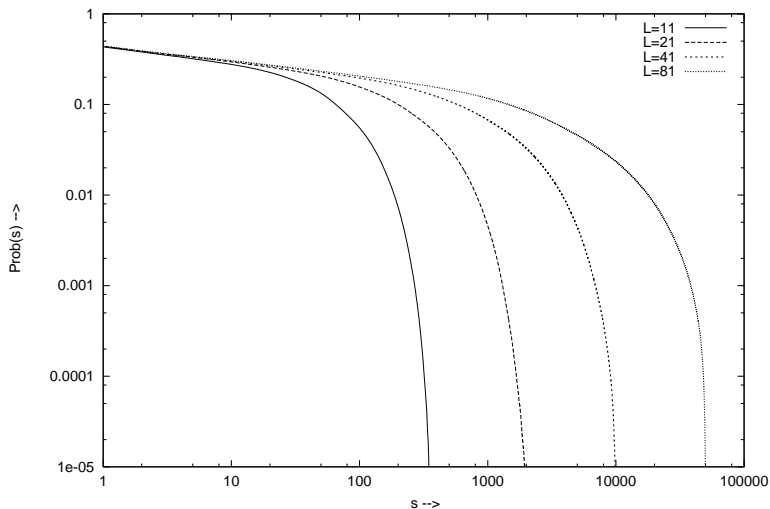}
\label{distsch} 
\caption{The cumulative probability ${\sf Prob}(s)$ that an avalanche is of
size greater than or equal to $s$, as a function of $s$ for different 
lattice sizes. The curves were generated by simulation with over $10^6$
events for each size of the lattice.}
\end{center}
\end{figure}

For readers who may like to write their own simulation program, we add
some tips here: to get good statistics, one needs to minimize the
finite-size effects. Avalanches that start near the boundary have a
different statistics than those that start away from it (in bulk). In the
limit of large sizes, the contribution of boundary avalanches to total
goes to zero as $(1/L)$. In simulations, one gets much better convergence
if the avalanches that start near the boundary are excluded from the data.
Else, the distributions for different sizes converge very slowly with $L$
even for small avalanche sizes. In Fig. 3, I used a cylinder of size $L
\times L$, with periodic boundary conditions in the $y$-direction, and open
boundary conditions in the x-direction. Particles were not added with
uniform probability everywhere. The odd-numbered particles were added only
along the middle ring away from the boundaries, with all sites on the ring
equiprobable. Even-numbered particles were added anywhere on the lattice.
Only avalanches resulting from the odd-numbered particles were used for
determining the avalanche distributions.  [We shall show later that this
does not affect the answer.] Imposing cylindrical boundary conditions
avoids corners, and consequent corrections to bulk behavior.

It is easy to see that in this model, events involving many topplings must
occur with significant probability. Since every particle added anywhere in
the system finally leaves the system, and on the average it must take at
least order $L$ steps to reach the boundary, we see that the average
number of topplings per added particle is at least of order $L$. In other
words,

\begin{equation}
\left<s\right> \equiv \sum s {\sf Prob}(s)  \ge c L,\ \mbox{for $L$
large.}
\end{equation}
where $c$ is some constant \cite{footnote2}. For large $L$, this tends to
infinity. Such a diverging expectation value can not be generated by any
distribution for which the probability of avalanches of size {\it larger
than} $s$ decreases with $s$ faster than $s^{-1-\epsilon}$, where
$\epsilon$ is any positive power. Thus the distribution must be of the
type sketched in fig. 3 with a small enough decay exponent.

This argument suggests that so long as sand can leave the system only at
the boundaries, the distribution of avalanches must have a power law tail.
In fact, careful experiments with real sand do NOT see such a tail.  How
to reconcile this with the rather robust argument given above?

It seems that the hypothesis of a single angle of repose is not quite
correct. There are at least two angles: an angle of stability $\theta_1$,
and an angle of drainage $\theta_2$, with $\theta_1$ greater than $
\theta_2$ by about a couple of degrees. As we add sand, the slope slowly
increases from $\theta_2$ to $\theta_1$, and then there is a single big
avalanche and the net slope decreases to $\theta_2$. In such a ``charge
and fire'' behavior, the big avalanches are nearly periodic.

Speaking mathematically, one can have diverging first moment of a
distribution with no power law tail, if the distribution consists of two
parts: a small events part which has almost all the weight of the
distribution, and a small weight (of order $L^{-a}$) at a rather large
value of $s \sim L^b$. As one needs to build up $O(L^3)$ particles to
increase the slope by a finite amount, and in a big event, a particle will
move by an amount of order $L$, this corresponds to $a = 3, b =4$.  Of
course, the existence of large events does not {\it exclude} a weak power
law tail in the small events.

Whether real granular media show SOC or not seems to depend on the shape
of grains. The scenario outlined above seems to describe sandpiles with
roughly spherical grains. However, if the grains are very rough and of
large aspect ratio, they behave differently. In a beautiful series of
experiments, Frette et al have shown that piles formed of long-grain rice
do show SOC behavior \cite{ricepile}.
 
The BTW model is not a very good model of real sand. For realistic
modelling of sand, the instability condition should be about a maximum
value of the slope, not a maximum value of the height. Such critical slope
models are easily defined, but not very tractable analytically. As a
pedagogical model to illustrate the basic ideas of SOC, the critical
height model is much better.  The bounded height variables in the BTW
model may be better viewed as fluctuations about the space-dependent mean
height of the pile. The analytical tractability of the model has led
to a lot of interest in the model, and  by now there is  a fairly large
amount of work dealing with this model. For recent reviews, see  
\cite{reviews}.

\section{The abelian sandpile model}
\subsection{Definition}

The BTW model has an important abelian property that simplifies its
analysis considerably. To bring out this abelian property in its full
generality, it is preferable to work with a generalized sandpile model to
be called the Abelian Sandpile Model (ASM) defined as follows~\cite{dhar90}:
  
We consider the model defined on a graph with $N$ sites labeled by
integers $1,2,\ldots N$ [Fig.4]. At each site $i$, the height of the
sandpile is given by a positive integer $z_i$.  At each time step, a site
is chosen at random from a given distribution, the probability of site $i$
being chosen being $p_i$. Its height is increased by 1 from $z_i$ to $
z_i+1$.

We are also given an integer $N \times N$ matrix $\Delta$, and a set of
$N$ integers $z_{i,c}$, $i=1$ to $N$.  If for any site $i$, $z_i>z_{i,c}$
then the site is said to be unstable, and it topples. On toppling at site
$i$, all heights ${z_j}$ are updated according to the rule

\begin{figure} 
\begin{center} 
\leavevmode
\psfig{figure=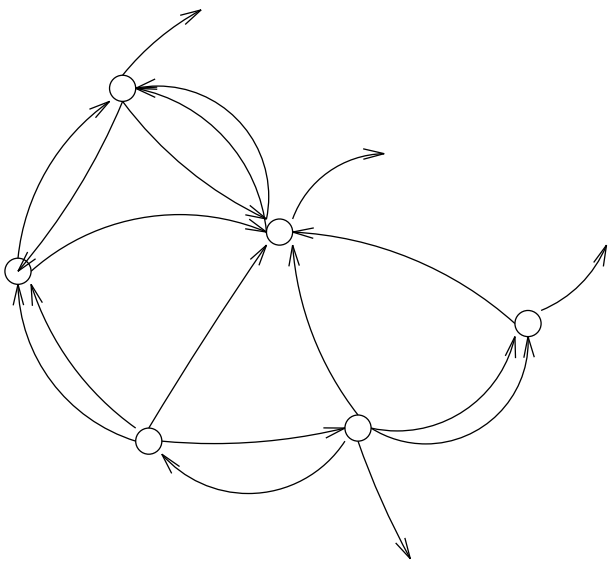,width=8cm,angle=0} 
\caption{ A graphical representation of a general ASM. Each node denotes a
site.  The maximum stable height $z_{i,c}$ at the site $i$ equals the
number of arrows going out of $i$. On toppling at any site, one particle
is transferred along each bond directed away from the site. } 
\end{center} 
\end{figure}

\begin {equation}
\mbox{If $z_i > z_{i,c}$, then $z_j\rightarrow z_j-\Delta_{ij}$, for every
$j$}. 
\end{equation}

Without loss of generality we may choose $z_{i,c} = \Delta_{ii}$ (this
amounts to a particular choice of the origin of the $z_i$ variables). In
this case, the allowed values of $z_i$ in a stable configuration are $1,2,
\ldots, \Delta_{ii}$.

Evidently the matrix $\Delta$ has to satisfy some conditions to
ensure that the model is well behaved. 

\begin{enumerate}

\item $\Delta_{ii}>0$, for every $i$. (Otherwise topplings never
terminate.)
\item For every pair $i\ne j$, $\Delta_{ij}\le0$. (This condition is
required to establish the Abelian property, as will be shown shortly.)
\item $\sum_j \Delta_{ij} \ge 0$ for every $i$. (This condition states
that sand is not generated in the toppling process.)
\item There is at least one site $i$ such that $\sum_j \Delta_{ij} > 0$.
Such sites are called dissipative sites. In addition, the matrix $\Delta$
is such that in the corresponding directed graph, there is a path of
directed links from any site to one of the dissipative sites. This
condition ensures that all avalanches terminate in a finite time.

\end{enumerate}

\subsection{ The Abelian Property}

Let $\cal C$ be a stable configuration, and define the operator $a_i$ such
that the stable configuration ${\cal C}'=a_i{\cal C}$ is the one achieved
after addition of sand at site $i$ and relaxing. The mathematical
treatment of the sandpile models relies on one simple property they
possess \cite{dhar90}: The order in which the operations of particle
addition and site toppling are performed does not matter. Thus the
operators $a_i$ commute, {\it i.e.},

\begin{equation}
 a_i a_j = a_j a_i\ ,\qquad \mbox{for every $i$, $j$.}
\end{equation}

To prove this we start by noting that if we have a configuration with two
or more unstable sites, then these sites can be relaxed in any order, and
the resulting configuration is independent of the order of toppling.
Consider two unstable sites $i$ and $j$.  If we topple at $i$ first, this
can only increase the value of $z_j$ (by condition 2), and site $j$
remains unstable. After toppling at $j$ also, the height at any site $k$,
as a result of these two topplings, undergoes a net change of
$-\Delta_{ik} - \Delta_{jk}$. Clearly toppling first at $j$, then at $i$
gives the same result. By a repeated use of this property, any number of
unstable sites in a configuration can be relaxed in any order, always
giving the same result.

Also, the operation of toppling at an {\it unstable} site $i$, commutes
with that of adding a particle at some site $j$, for if the site $i$ is
unstable and topples, it will do so also if the addition at site $j$ was
performed first.  By a repeated use of the property that the individual
operations of addition and toppling commute with each other, the abelian
property of $\{a_i\}$ follows.

While this property seems very general, it should be noted that it is not
shared by most of the other models of SOC like the forest-fire model
\cite{forestfire}, or even other sandpile models, such as the critical
slope model where the toppling condition at site $i$ depends on the height
at other sites, or the Zhang model \cite{zhang} in which the amount of
sand transferred depends on the amount by which the local height exceeds
the critical value.

\subsection{ Operator Algebra}

The operators $\{a_i\}$ satisfy additional relations. For example, on the
square lattice, if one adds 4 particles at a given site, the site is bound
to topple once, and a particle is added to each of its nearest neighbors.
Thus,

\begin{equation}\label{gen4}
a_j^4=a_{j_1}a_{j_2}a_{j_3}a_{j_4}\ ,
\end{equation}
where $j_1,\ldots,j_4$ are the nearest neighbors of $j$ [Fig. 5].
\begin{figure}
\begin{center}
\leavevmode
\psfig{figure=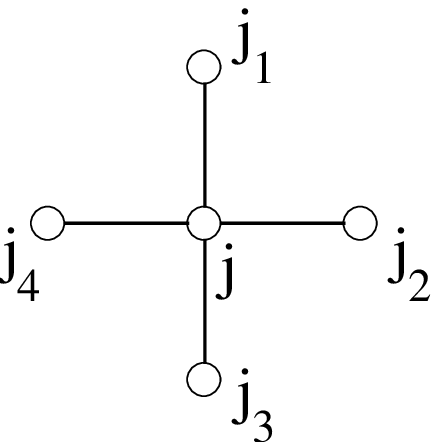,width=4cm,angle=0}
\caption{ }
\end{center}
\end{figure}

In the general case one has instead of (\ref{gen4}),
\begin{equation} \label{gen}
a_i^{\Delta_{ii}}=\prod_{j\ne i}a_j^{-\Delta_{ij}}\ .
\end{equation}

Using the abelian property, in any product of operators $\{a_i\}$, we can
collect together occurences of the same operator, and using the reduction
rules (\ref{gen}), we can reduce the power of $a_i$ to be always less than
$\Delta_{ii}$. The $\{a_i\}$ are therefore the generators of a finite
abelian semi-group (the associative property follows from the definition),
subject to the relations (\ref{gen}). These relations define the
semi-group completely.

Let us consider the repeated action of some generator $a_1$ on some
configuration $\cal C$. Since the number of possible states is finite, the
orbit of $a_1$ must at some stage close on itself, so that $a_1^{n+p}{\cal
C}=a_1^n{\cal C}$ for some positive period $p$, and non-negative integer
$n$. The first configuration that occurs twice in the orbit of $a_1$ is
not necessarily ${\cal C}$, so that the orbit consists of a sequence of
transient configurations followed be a cycle (see fig. 6). If this orbit
does not exhaust all configurations, we can take a configuration outside
this orbit, and repeat the process. Thus the space of configurations is
broken up into disconnected parts, each containing one limit cycle.

\begin{figure}
\begin{center}
\epsfbox{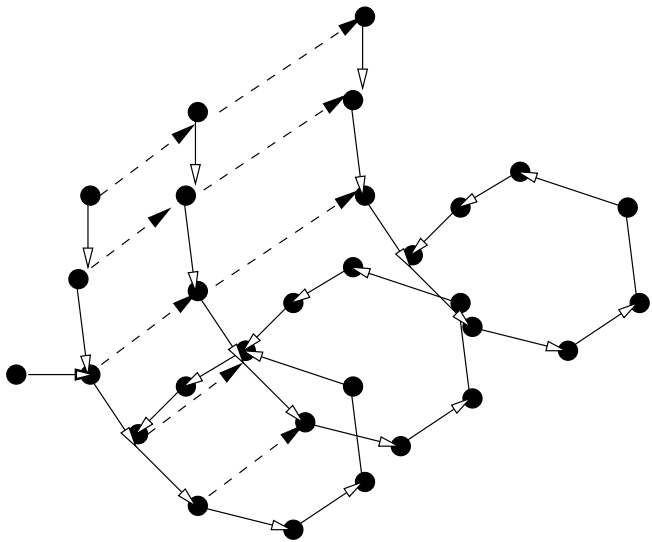}
\label{cycle}
\end{center}
\caption{A schematic representation of action of  addition operators
on the stable configurations. Each point represents a configuration. 
Arrows with full lines denote action of the operator $a_1$. The arrows
with dotted lines denote action of the operator $a_2$.}
\end{figure}

Under the action of $a_1$ the transient configurations are unattainable
once the system has reached one of the periodic configurations. In
principle these states might still be reachable as a result of the action
of some other operator, e.g.  $a_2$. However, the abelian property implies
that if ${\cal C}$ is a configuration which is part of one of the 
limit cycles
of $a_1$, then so is $a_2{\cal C}$, since $a_1^p{\cal C}={\cal C}$ implies
that $a_1^pa_2{\cal C}=a_2a_1^p{\cal C}=a_2{\cal C}$. Thus, the transient
configurations with respect to $a_1$ are also transient with respect to
other operators $a_2, a_3, \ldots$ , and hence occur with zero probability
in the steady state of the system. The abelian property thus implies that
$a_2$ maps cycles of $a_1$ to cycles of $a_1$, and moreover, that all
these cycles have the same period.

We can now repeat our previous argument, to show that the action of $a_2$
on a cycle is finally closed on itself to yield a torus, possibly with
some transient configuration, which may be also discarded. Repeating this
argument for the other generators leads to the conclusion that the set of
all recurrent configurations forms a multi-dimensional torus under the
action of the $a$'s.

\subsection{The Abelian Group}

If we restrict ourselves to the set $\bf R$ of recurrent states, we can
define inverse operators $a_i^{-1}$, for all $i$ as each configuration in
a cycle has exactly one incoming arrow corresponding to operator $a_i$.
Thus the $a$ operators generate a group.  The action of the $a$'s on the
states corresponds to translations on the torus. From the symmetry of the
torus under translations, it is clear that all recurrent states appear in
the stationary state with equal probability.

This analysis, which is valid for every finite abelian group, leaves open
the possibility that some configurations are not reachable from each
other, in which case there might be several different mutually
disconnected tori. However, such a situation cannot happen in the ASM if
we allow addition of sand at all sites with nonzero probabilities (all
$p_i > 0$). Then, the configuration ${\cal C}_{\max}$, in which all the
sites have their maximal height is reachable from every configuration, and
is therefore recurrent, and since inverses of  $a$'s exist for
configurations in $\bf R$, every
recurrent configuration is reachable from ${\cal C}_{\max}$, implying that
all recurrent states lie on the same torus.

If some of the addition probabilities $p_i$ are zero, one has to consider
the subgroup formed taking various powers of those $a_j$'s which have
nonzero $p_j$. This may still generate the full group. (On a square
lattice of size $L \times L$, the number of independent generators of the
group can be shown to be only $L$ \cite{drsv95}, and hence if one selects
$L'>L$ generators at random, with a large probability one generates the
full group.) If not, the dynamics splits ${\bf R}$ to several disconnected
tori, and the steady state is non-unique. It does not fully forget the
initial condition, as the evolution is confined to the torus on which we
started.

\subsection{ The Evolution Operator and the Steady State}

We consider a vector space $\cal V$ whose basis vectors are the different
configurations of $\bf R$. The state of the system at time $t$ will be
given by a vector 
\begin{equation}
\left|P(t)\right>\ =  \sum_{\cal C} {\sf Prob}({\cal C},t)\left|{\cal
C}\right>\ ,
\end{equation}
where ${\sf Prob}({\cal C},t)$ is the probability that the system is in
configuration ${\cal C}$ at time $t$. The operators ${a_i}$ can be defined
to act on the vector space ${\cal V}$ through their operation on the basis
vectors. 

The time evolution is Markovian, and governed by the equation
\begin{equation}
\left|P(t+1)\right> = {\cal W } \left|P(t)\right>
\end{equation}
where
\begin{equation}
{\cal W} = \sum_{i=1}^N  p_i a_i
\end{equation}

To solve the time evolution in general, we have to diagonalize the
evolution operator ${\cal W}$.  Being mutually commuting, the $a_i$ may be
simultaneously diagonalized, and this also diagonalizes ${\cal W}$. Let
$\left|\{\phi\}\right>$ be the simultaneous eigenvector of $\{a_i\}$, with
eigenvalues $\{e^{i\phi_i}\}$, for $i = 1,\ldots N$. Then
\begin{equation}  \label{eigen}
a_i\big|\{\phi\}\big>=e^{i\phi_i}\big|\{\phi\}\big>  
{\rm ~~~for} ~~i~=~1 {\rm ~to}~~ N. 
\end{equation}
Since the $a$ operators now form a group, the relations (\ref{gen})
may be written as

\begin{equation} \label{ggen}
\prod_j a_{j}^{\Delta_{kj}}=1\ ,\qquad\mbox{for every $k$.} 
\end{equation}
Applying the LHS to the eigenvector $\left|\{\phi\}\right>$ gives
$\exp(i\sum_j\Delta_{kj}\phi_j)=1$, for every $k$, so that
$\sum_j\Delta_{kj}\phi_j=2\pi m_k$, or inverting,

\begin{equation} 
\phi_j=2\pi\sum_k\left[\Delta^{-1}\right]_{jk}m_k\ ,
\end{equation}
where $\Delta^{-1}$ is the inverse of $\Delta$, and the $m_k$'s are
arbitrary integers.

The particular eigenstate $\left|\{0\}\right>$ ($\phi_j=0$ for all $j$)
is
invariant
under the action of all the $a$'s,
$a_i\left|\{0\}\right>=\left|\{0\}\right>$. Thus
$\left|\{0\}\right>$ must be the stationary state of the system since

\begin{equation} 
\sum_i p_i a_i \left|\{0\}\right>= \sum_i p_i \left|\{0\}\right>=
\left|\{0\}\right>\ .
\end{equation}

We can now see explicitly that the steady state is independent of the
values of the $p_i$'s and that {\it in the steady state, all recurrent
configurations occur with equal probability.}. As discussed above, when
some of the $p_i$'s are zero, the stationary state may not be unique. In
this case, the general steady state is a linear combination of two or more
eigenvectors of ${\cal W}$ having eigenvalue $1$ and then the dynamics is
no longer ergodic.

\subsection{Propagation of Avalanches : The Two-point Function}

An important quantity characterizing the steady state is its response to a
small perturbation. Here it can be measured by the probability that a
single particle added at some site causes a perturbation which affects a
site at a distance $R$ from it. Let $G_{ij}$ be the expected number of
toppling at site $j$, upon adding a particle at site $i$ {\it in the
steady state}. Since in the steady state the expectation value of the
number of particles at a site is constant, there must exist a balance
between the average influx of particles from toppling at other sites, and
the outflux from toppling at the site. For the square lattice model, this
relation gives [Fig. 5]

\begin{equation} 
G_{ij_1}+G_{ij_2}+G_{ij_3}+G_{ij_4}=4G_{ij}\, 
\end{equation}
where   $i \neq j$. In the general case we have

\begin{equation}\label{G} 
\sum_k G_{ik}\Delta_{kj}=\delta_{ij}\ ,
\end{equation}
where the $\delta_{ij}$ takes account of the addition of one particle
at $i$. Solving for $G$ gives simply

\begin{equation} 
G_{ik}=\left[\Delta^{-1}\right]_{ik}\ . 
\end{equation}
Note that this derivation does not use the abelian property, only the fact
that the mass transferred in each toppling is the same. It is therefore
valid for a much wider class of models. Such an equation still holds if
the toppling condition is a critical slope condition, but not if the
number of particles transferred depends on the slope.

In the case of a $d$-dimensional hypercubic lattice, with particle trasfer
to nearest neighbors, and loss of sand only at the boundary, this is just
the inverse of the lattice Laplacian. Then it is easy to see that

\begin{eqnarray}
G_{ik} &\sim& r_{ik}^{2-d},\qquad \mbox{for  $d > 2$};\\
       &\sim& \log(L/r_{ik}), \qquad \mbox { for $d = 2$};\\
       & \sim& (L-r_{ik}),\qquad \mbox{ for d = 1}.\end{eqnarray}
where $L$ is the size of the lattice. We note that  the influence
function G does show the promised long-ranged correlations.

\section{ Recurrent and Transient Configurations}

Given a stable configuration of the pile, how can we tell if it is
recurrent or transient?  A first observation is that there are some
forbidden subconfigurations which may never be created, if not already
present in the initial state by addition of sand and relaxation. The
simplest example on the square lattice case is a subconfiguration of two
adjacent sites of height 1,
\mbox{\begin{tabular}{|c|c|}\hline 1&1\\\hline\end{tabular}}~. Since
$z_i>0$, a site of height 1 may only be created as a result of a toppling
at one of the two sites (toppling anywhere else can only add particles to
this pair of sites). But a toppling of either of these sites results in a
height of at least 2 at the other. Thus, any configuration which contains
two adjacent 1's is transient. The argument may now be repeated to prove
that subconfigurations such as
$$\mbox{\begin{tabular}{|c|c|c|} \hline1&2&1\\\hline\end{tabular}}\ ,\qquad 
\mbox{\begin{tabular}{|c|c|}\hline 1&2\\\hline\multicolumn{1}{c|}{}
&1\\\cline{2-2}
\end{tabular}}\ ,\qquad  
\mbox{\begin{tabular}{|c|c|c|} \cline{2-2} \multicolumn{1}{c|}{}&1&
\multicolumn{1}{|c}{}\\\hline1&4&1\\\hline\multicolumn{1}{c|}{}
&1&\multicolumn{1}{|c}{}\\\cline{2-2}
\end{tabular}}\  $$ 
also can never appear in a recurrent configuration.

In general, a forbidden subconfiguration (FSC) is a set of connected sites
$F$ such that the height $z_j$ of each site $j$ in $F$, is less than or
equal to the number of neighbors of $j$ in $F$. The proof of this
assertion is by induction on the number of sites in $F$. For example,
creation of the \mbox{\begin{tabular}{|c|c|c|}\hline
1&2&1\\\hline\end{tabular}} configuration, must involve toppling at one of
the end sites, but then the configuration must have had a
\mbox{\begin{tabular}{|c|c|}\hline 1&1\\\hline\end{tabular}} before the
toppling. But this was shown before to be forbidden, etc.

\subsection{ The `Multiplication by Identity'  Test}

A more systematic way to test for recurrence of a given configuration is
by using the relations~(\ref{gen}). Let us take up again the example of
square lattice. Multiplying all the relations gives
\begin{equation} \prod_i a_i^4=\prod_i a_i^{n_i}\ ,\end{equation}
where $n_i$ is the number of neighbors of a given site, {\it i.e.}, 4
for a bulk site, 3 or 2 for a boundary site. Now we use the property
that inverses are defined on the set $\bf R$ of recurrent states,
which allows cancellation, yielding
\begin{equation} 
\prod_i a_i^{4-n_i}=1\ .
\end{equation}

In other words, in order to check whether a given configuration $\cal C$
is recurrent, one has to add a particle at each boundary site (2 at
corners), relax the system, and check whether the final  configuration
is same as $\cal C$. If it is, $\cal C$ is recurrent, otherwise it is not
\cite{speer}. 

An interesting consequence of the existence of forbidden configurations is
the following: Consider an ASM on an undirected graph with $N_s$ sites and
$N_b$ bonds between sites.  Then in any recurrent configuration the number
of sandgrains is greater than or equal to $N_s + N_b$. Note that here we
do not count the boundary bonds that correspond to particles leaving the
system. To prove this, we just observe that if the inequality is not true for
any configuration, it must have an FSC in it (proof by induction on
$N_s$).

\subsection{The Burning Test}

As explained above, a recurrent configuration is characterized by being
invariant under addition of particles at the boundary in a judicious
fashion, and relaxing. When the matrix $\Delta$ is symmetric, this test is
equivalent to a simpler test, called the burning test, which is defined as
follows \cite{md92}. Given a configuration, at first all the sites are
considered unburnt. Then, burn each site whose height is larger than the
number of its unburnt neighbors. This process is repeated recursively,
until no further sites can be burnt. Then, if all the sites have been
burnt, the original configuration was recurrent, whereas if some unburnt
sites were left, then the original configuration was transient, and the
remaining sites form an FSC.

As an example, let us take a configuration on a four by four square
lattice  shown below. Its evolution under  burning is as shown:
$$\mbox{\begin{tabular}{|c|c|c|c|} \hline
4 & 2 & 3 & 4 \\ \hline
1 & 4 & 2 & 2 \\ \hline
3 & 4 & 3 & 3 \\ \hline
4 & 1 & 3 & 2 \\ \hline \end{tabular}}
\longrightarrow
\mbox{\begin{tabular}{|c|c|c|c|} \hline
  & 2 & 3 &   \\ \hline
1 & 4 & 2 & 2 \\ \hline
3 & 4 & 3 & 3 \\ \hline
  & 1 & 3 & 2 \\ \hline \end{tabular}}
\longrightarrow
\mbox{\begin{tabular}{|c|c|c|c|} \hline
  & 2 &   &   \\ \hline
1 & 4 & 2 & 2 \\ \hline
  & 4 & 3 & 3 \\ \hline
  & 1 & 3 & 2 \\ \hline \end{tabular}}
\longrightarrow
\mbox{\begin{tabular}{|c|c|c|c|} \hline
\ &   &   &   \\ \hline
  &   & 2 & 2 \\ \hline
  &   & 3 & 3 \\ \hline
  & 1 & 3 & 2 \\ \hline \end{tabular}}\ ,
$$
where burned sites are denoted by empty squares. At first the `fire' 
burns some of the  boundary sites, and it then advances inward. No
site in the last
configuration can be burned, and the set of unburnt sites is thus a
forbidden subconfiguration. If the height of the third site on the second
row were changed from 2 to~3, then it could be burned next, and the fire
would have spread completely, and such a configuration would be recurrent.

It is obvious that the burning process is in fact identical to toppling,
starting from addition at the boundary, except that it is simpler to use,
as we work with initial height configuration, and do not need to keep
updating them with each toppling [since we know that the final
configuration is the same as the starting one, if all sites topple exactly
once].

To see what could go wrong when $\Delta$ is asymmetric consider the
example of a two site sandpile with a toppling matrix
$\left(\begin{array}{rr}7&-4\\-2&2\end{array}\right)$. In the
unsymmetrical case, a site is burnt if its height exceeds the number of
outgoing bonds to unburnt neighbors.  A configuration with five grains at
1, and one grain at 2 would pass the burning test, but is in fact
transient. A site like site 2, which has more incoming arrows than
outgoing arrows is called a greedy site. In this case, equations satisfied
by the operators $a_1$ and $a_2$ are $a_1^7 = a_2 ^4$, and $a_2^2 = a_1
^2$. This gives the identity operator as $a_1^3 = 1$, but adding three
particles on site $1$ in a recurrent configuration, we get one toppling at
site $1$, but two topplings at site $2$. Clearly, the burning test fails
as it assumes that under multiplication by the identity operator
[Eq.(11)], each site topples only once. We also see that if there are no
greedy sites, the burning test is a necessary and sufficient test for
recurrence even for unsymmetrical graphs.

\subsection{Number of Recurrent Configurations}

What is the total number of recurrent configurations? If we were to take
care of only the constraint that \mbox{\begin{tabular}{|c|c|}\hline
1&1\\\hline\end{tabular}} is disallowed, this problem is a special case of 
the nearest-neighbor exclusion lattice gas model, which is equivalent to
the famous unsolved problem of the Ising model in a magnetic field.
However, it turns out the infinity of the FSC constraints actually can
be taken care of, and the final result is rather simple: the total number
of recurrent configurations $|{\bf R}|$ is given by 
\begin{equation}\label{det} |{ \bf R}| = \det \Delta\ .
\end{equation}

Consider the set of all possible configurations
$\{z_i\},\,-\infty<z_i<\infty$. We can define an equivalence
relation on this set of configurations as follows: We call two
configurations ${\cal C}$ and ${\cal C'}$ as equivalent if one can be
obtained from the other by a series of topplings and untopplings
[untoppling is the reverse of toppling, in an 
`untoppling' at site $i$, the height at all sites $j$ is {\it increased}
by an amount $\Delta_{ij}$]. For this purpose, we can topple or untopple
at 
any site whatever be its local height at that time. This is equivalent to
defining  
${\cal C}\sim{\cal C}'$ if and only if
\begin{equation}
z_j=z_j'-\sum_i m_i\Delta_{ij}
\end{equation}
for some integers $\{m_i\}, i=1,\ldots,N$. The members of an
equivalence class form a superlattice of ${\bf Z}^N$, with the rows of
$\Delta$ as basis vectors [Fig. 7]. It is easily shown that each
equivalence class
contains exactly one recurrent configuration. 

\begin{figure}
\begin{center}
\psfig{figure=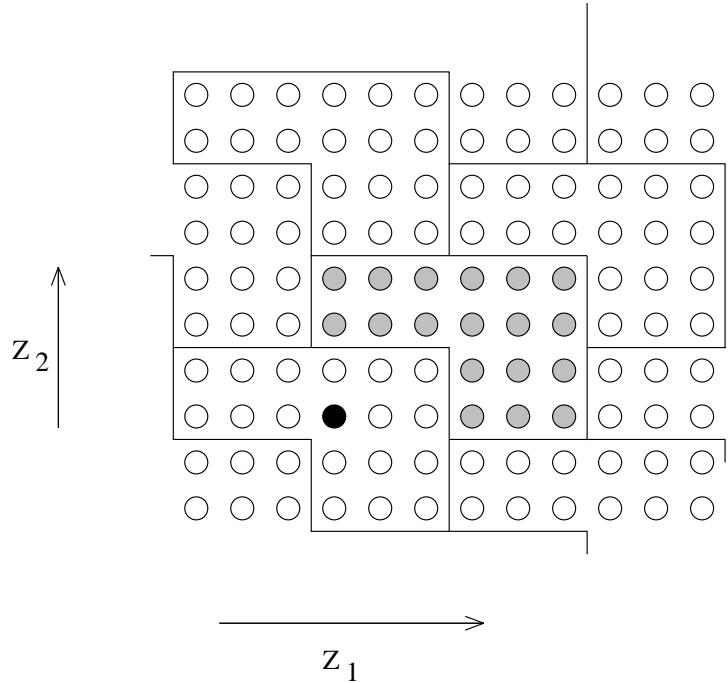,width=9cm,angle=-90}
\caption{The superlattice of equivalent configurations in the space of  
of all configurations
$(z_1,z_2)$ for a 2-site automata with $\Delta =
\left[\protect\begin{array}{rr} 6 & -2 \protect\\ -3 & 4
\protect\end{array}\right]$ with copies of the recurrent set $\cal R$. The
grey vertices are the recurrent
configurations.  The black circle marks the configuration $(0,0)$. }
\end{center}
\end{figure}

Firstly, starting with any configuration, if we untopple once at all
sites, we get an equivalent configuration with an increased number of
sandgrains in the system. We then relax the resulting configuration.  As
this process, equivalent to adding particles and relaxing, can be
repeated, the stable configuration thus reached is recurrent.  Thus, each
equivalence class has at least one recurrent configuration.

Now, we argue that there can be at most one recurrent configuration in
each equivalence class. Assume the contrary. Let $\{z_i\}$ and $\{z_i'\}$
be two equivalent stable configurations. Let $m_{max}$ be the maximum
value of $m_i$ in Eq. (23) over all $i$, and $F$ be the set of sites for
which this maximum value is reached. Then, to go from $\{z_i'\}$ to
$\{z_i\}$, each site in $F$ topples at least one more time than its
neighbors outside $F$. Hence each site in $F$ would lose one particle to
each neighbor outside $F$ in going from $\{z'_i\}$ to $\{z_i\}$. If
$\{z_i'\}$ is stable, this implies that $F$ must be an FSC in $\{z_i\}$.
This implies that $\{z_i\}$ is not recurrent, which contradicts the
assumption.  Therefore, the number of elements of $\bf R$ must be equal to
the volume of the unit cell of the superlattice. But this is clearly $\det
\Delta$. This proves Eq.(\ref{det}).

\section{Algebraic Structure of the Abelian Group}

Any finite abelian group $Z$ can always be expressed as a product of
cyclic groups in the following form:
\begin{equation} Z={\bf Z}_{d_1}\times\cdots\times{\bf Z}_{d_g}\ .
\label{can}\end{equation}
That is, the group $Z$ is the direct product of $g$ cyclic groups of
orders $d_1,\ldots,d_g$. Furthermore the integers $d_1\ge d_2\ge \cdots
d_g>1$, can be chosen such that $d_i$ is an integer multiple of $d_{i+1}$,
and under this condition, the decomposition is unique. For example, ${\bf
Z}_{6}\times{\bf Z}_{4}\cong{\bf Z}_{12}\times{\bf Z}_2$ but ${\bf
Z}_{2}\times{\bf Z}_{2}\not\cong{\bf Z}_{4}$.

To construct this factorization for the abelian group $G$ for the ASM, we
use the following classical results, known as the Smith decomposition
\cite{smithdecom}: Given an integer $N\times N$ matrix $\Delta$, there
exist unimodular integer $N\times N$ matrices $A$ and $B$, and an integer
diagonal matrix $D$ such that
\begin{equation} \Delta=ADB\ .\end{equation}
[A matrix is said to be unimodular if its determinant is $\pm 1$.  The
inverse of an integer unimodular matrix is also an integer matrix.] The
matrix $D$ is unique if we further impose the condition that its diagonal
elements $d_1,\ldots,d_N$ are such that $d_i$ is a multiple of $d_{i+1}$.
The matrices $A$ and $B$ are not unique.

The Smith decomposition is constructed step by step as follows: We define
two matrices $\Delta$ and $\Delta'$ to be similar [to be denoted by
$\Delta\sim\Delta'$] if there exist unimodular matrices $S_1$ and $S_2$
such that $\Delta=S_1\Delta'S_2$.  Among the unimodular matrices there are
matrices $P_{ij}$ which interchange the rows $i$ and $j$ acting on the
left and columns $i$ and $j$ acting on the right, as well as $Q_{ij}(n)$
which add $n$ times row (column) $j$ to row (column) $i$ acting on the
left (right). Explicitly, $P_{ij}$ is just the identity with rows $i$ and
$j$ interchanged, and $Q_{ij}(n)$ is the identity with an additional $n$
in the $(ij)$-th position.

Let $\Delta_{ij}$ be one of the non-zero elements of $\Delta$ having the
smallest absolute value. Operating with interchange matrices, we can get
an equivalent matrix $\Delta'$ in which this element occurs in the lower
right corner. Then, by multiplication with the matrices of type $Q$, the
members of the last row and last column are replaced by their remainders
modulo $m$. If any of the remainders are non-zero, the smallest one may be
shifted to the corner, and the process repeated until the last row and
column are zero except for the corner element:
\begin{equation} \Delta\sim
\left(\begin{array}{cc}\Delta^{(1)}&\begin{array}{c}0\\\vdots\end{array}
\\ 0\,\cdots&d_N\end{array}\right)
\ .\end{equation}

 Then, if any of the still nonzero elements of the resulting matrix is not
divisible by the element $d_N$ in the lower right corner, its row may be
added to the last row, and as before the remainder is brought to the
corner. Thus, $d_N$ can be chosen to be the greatest common divisor of all
the elements of the original matrix $\Delta$.

Applying the same process reduces $\Delta^{(1)}$ to matrix $\Delta^{(2)}$
which has additionally row and column $N-1$ empty except for the
$(N-1,N-1)$-th element which is the gcd of the elements of $\Delta^{(1)}$
except $d_N$. This procedure finally yields the matrix $D$ and the
products of $P$ and $Q$ matrices used give the matrices $A$ and $B$.
 
Given the Smith decomposition of $\Delta$, we can define generators of the
cyclic subgroups of the group $G$ by
\begin{equation} 
 e_\alpha= \prod_ja_j^{B_{\alpha j}},\quad  \alpha=1,\ldots,g\ .
\end{equation}
It is then easy to see that 
\begin{equation}
e_\alpha^{d_\alpha} = \prod_j a_j^{[DB]_{\alpha j}} = \prod_j
a_j^{[A^{-1}\Delta]_{\alpha j}} = 1,\quad \alpha=1,\ldots, g\ .
\end{equation}
where we have used Eq. [11].

For any configuration $\{z_i\}$, we define scalar functions
$I_\alpha$, for $\alpha = 1\ldots g$, linear in the
height variables by the equations
\begin{equation}
I_\alpha=\sum_{j=1}^Nz_j\left[B^{-1}\right]_{j\alpha} ,\quad
\end{equation}

As a result of toppling at $k$, the change in $z_j$ is by an amount
$-\Delta_{kj}$, so that the change in $I_\alpha$, denoted by $\delta
I_\alpha$, is
\begin{equation} 
\delta I_\alpha=-\sum_j\Delta_{kj}\left[B^{-1}\right]_{j\alpha}=
-\left[\Delta B^{-1}\right]_{k\alpha}=-[AD]_{k\alpha}=-A_{k\alpha}d_\alpha
\ ,
\end{equation}
which shows that $I_\alpha\bmod d_\alpha$ is invariant under toppling.
Thus, it follows that with each equivalence class is associated a single
set of integers $\left[I_1,\ldots,I_g\right]$, where $I_\alpha$ is defined
modulo $d_\alpha$.

The action of the operators $e_\alpha$ on the $I$'s is such that
\begin{equation} 
I_\beta\left(e_\alpha\{z\}\right)-I_\beta(\{z\})=
\sum_j B_{\alpha j}\left[B^{-1}\right]_{j\beta}=\delta_{\alpha\beta},\ 
\label{actione}
\end{equation}
which means that the $e_\alpha$'s are the generators of the cyclic
subgroups in the decomposition (\ref{can}), and that the $I_\alpha$'s
can serve as coordinates along the respective cycles.

Eq.~(\ref{actione}) also implies that each of the possible
values of $\left[I_1,\ldots,I_g\right]$ is reachable. Since the
number of such configurations is $\prod_{\alpha=1}^g d_\alpha=\det\Delta$, 
we conclude that each recurrent configuration may be uniquely labeled
by the $\{I\}$ variables.

Calculation of properties in the steady state involves
an average over all recurrent states. Using the height variables
to characterize configurations is not very convenient here, because the
existence of FSC's implies many nontrivial constraints in the values of
$\{z_i\}$ used in the sum. The summations over $I_\alpha$ are independent
of each other ,
hence formally simpler. The explicit calculation of the matrices $A$
and $B$ is however nontrivial, and has  not been done in any nontrivial
case.
 
The toppling invariants $\{I\}$ are very useful in another way. Consider 
a general toppling invariant $I$ which is linear in the height
variables $\{z_i\}$, of the form
\begin{equation}
I_{\alpha}= \sum_{i=1}^N g_{i\alpha} z_i ~~~~~(mod ~~d_{\alpha})
\end{equation} 
where $g_{i\alpha}$ are some integer constants. The values of the
constants depends on $\Delta$.
We define unitary operators $S_{\alpha}$ corresponding the toppling
invariant $I_{\alpha}$ by the equation
\begin{equation}
S_{\alpha} = \exp(2 \pi i I_{\alpha} / d_{\alpha}), ~~\alpha= 1,\ldots g.
\end{equation}
These operators are diagonal operators in the configuration basis. It is
easy to see that their commutation with $\{a_i\}$ are given by the
equation
\begin{equation}
S_{\alpha} a_i S_{\alpha}^{-1} =  \exp(2\pi i g_{i\alpha}/d_{\alpha}) a_i
\end{equation}
This implies that if $|\psi\rangle$ is a simultaneous right eigenvector of the
operators $\{a_i\}$ with eigenvalues $\{\rm a_i\}$, then $S|\psi\rangle$ 
is also
a right eigenvector of these operators with a different set of eigenvalues
$\exp(2 \pi i g_{i\alpha}/d_\alpha) {\rm a_i}$. Thus $S_\alpha$ act as
ladder operators also for the evolution operator ${\cal W}$.

\section{Spanning Trees, Resistor Networks, Potts Model and Loop-Erased
Random Walks}

In this section, we shall use the burning test to relate the  ASM
to the classical spanning tree problem. This then allows one to relate the
ASM problem to other well-studied problems in lattice statistics and graph
theory: the resistor networks, the Potts model, and the loop-erased random
walk.

Spanning trees are graph theoretical objects, defined as follows: Consider a
symmetric graph, where multiple links are allowed, but not self
links. A spanning tree is defined to be a connected set of edges which
touches all the nodes, but contains no loops (see fig. 8).

\begin{figure}
\begin{center}
\center{figure}
\epsfbox{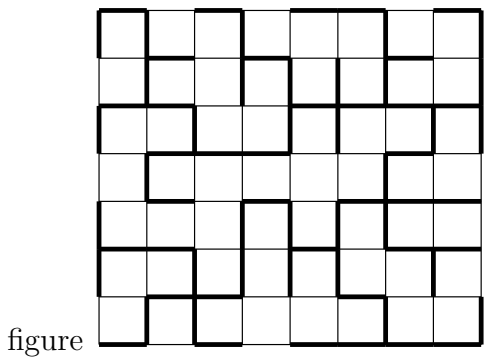}
\label{tree} 
\end{center}
\caption{A spanning tree (marked by thick lines) on a $9 \times 8$
square lattice .}
\end{figure}

There is a symmetric matrix $M$ associated with each such symmetric graph,
such that the off-diagonal element $M_{ij}$ is the negative of the number of
links between nodes $i$ and $j$, and $M_{ii}$ equals the number of links
attached to node $i$. The well-known matrix-tree theorem in graph theory
\cite{harary} states that the number of different spanning trees one can
construct on the graph is equal to the determinant of a matrix obtained by
deleting an arbitrary column and an arbitrary row from $M$.

\subsection{ Relation between Spanning Trees and ASM}

In order to relate the undirected  ASM to the spanning trees problem, we
consider the graph obtained by adding a sink site to the (undirected)
graph specifying the ASM, and connect the sink site to all the boundary
sites (called dissipative sites previously) by as many bonds as the
number of sandgrains dissipated at that site.  We now take any recurrent
configuration of the ASM, and consider the propagation of `fire' in the
burning test for this configuration.  The fire starts at the sink site
from which spreads to boundary sites, and then onto other sites. Since
each site is visited at most once, propagation of fire in the burning
process describes a tree, and for a recurrent configuration a spanning
tree.

We now show that one can uniquely determine the initial height
configuration if one knows  the spanning tree formed by the fire. This
would establish a one-to-one correpondence between the recurrent
configurations of the ASM and the spanning trees on the the corresponding
graph (with sink site added).

In the formulation of the burning test, the order in which various sites
are burnt is immaterial. To establish the equivalence, we make a
particular choice: The sites are burnt in discrete `time' steps. At
time-step $t=0$, only the sink site is assumed to be burnt, and all other
sites are unburnt. For all $t > 0$, at time $t+1$, we burn all sites which
are burnable at time $t$ (i.e. sites whose height exceeds the number of
unburnt neighbors). Thus at time $1$, only some of boundary sites are
burnt. At the next time-step, the fire will spread to some of their
neighbors, and so on. If a site catches fire at time $t$, it is said to
have a burning time $t$. Clearly, if the burning time of a site is $t$, it
must have at least one site with a burning time exactly $t-1$.  We form
the
spanning tree corresponding to this recurrent configuration by connecting
each site to its neighbor with burning time one less than its own.

It is easy to see that in this way, knowing the tree, we can determine the
the height at each site. For example, on a square lattice, an edge site
(not a corner site) has a burning time $1$ if and only if its height is
$4$. If a site $i$ with burning time $t_i$ has $r$ neighbors with lesser
burning time, its height must be $(5-r)$ [because it did not catch fire
until $r$-th neighbor was burnt].  There can be some ambiguity, if two or
more neighbors of $i$ have burning time exactly $(t_i-1)$. To remedy this,
we adopt some arbitrary precedence rule to decide the burning path
depending on the height at that site. If there are $s$ neighbors of the
site with burning time exactly $t_i-1$, there are exactly $s$ possible
values of $z_i$ consistent with the burning history, hence such a rule can
always be set up. For example, we may choose the rule that if the height
at the site could be $3$ or $4$ consistent with the burning rules, then
height $3$ would assign the burning path to the neighbor coming first
according to the precedence rule $N>E>S>W$, and if height is $4$ to the
other. In this way, we can set up a one-to-one correspondence between
recurrent configurations of the ASM and all possible spanning trees on the
corresponding graph (with sink site added).

If one constructs the matrix $M$ associated with the graph of the ASM,
deleting the row and column of the sink site, one recovers the toppling
matrix $\Delta$. By the matrix tree theorem, the number of spanning trees
on the graph is $\det\Delta$, and by the one-to-one correspondence, this
is also the number of recurrent configurations, as we have already seen.

\subsection{ Relation to the Resistor Network model}

The connection of spanning trees makes the ASM problem equivalent to
another classical problem in lattice statistics: the resistor network
problem. We are given a set of resistors with different conductances
connected to each other in some way. Suppose there are $N$ nodes with 
a resistor with conductance $\sigma_{ij}$ between the nodes $i$ and $j$.
[If there is no resistor beteen $i$ and $j$, we set $\sigma_{ij} =0$]. The
problem is to find the equivalent resistance between any two given nodes
of the network. 
\begin{figure}
\begin{center}
\leavevmode
\psfig{figure=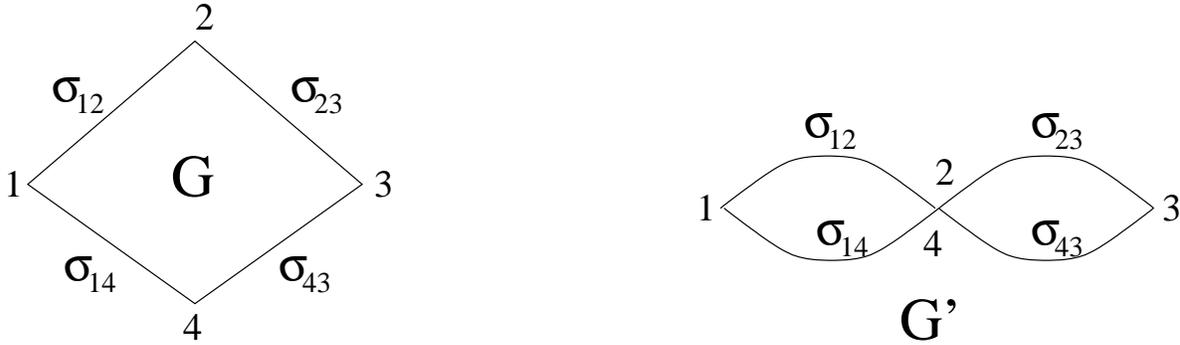}
\caption{ An example of the Kirchoff formula for resistor networks. The
coductances of individual links are shown in the network graph $G$.
Identifying the sites $2$ and $4$ we get the graph $G'$. In this case,
$T(G)=\sigma_{12} \sigma_{23} \sigma_{14} \sigma_{43}( \sigma_{12}^{-1}
+\sigma_{23}^{-1} +\sigma_{14}^{-1} +\sigma_{43}^{-1})$ and $T(G') =
(\sigma_{12} +\sigma_{14})(\sigma_{23} +\sigma_{43}$). Here $R_{24} =
T(G')/T(G)$}
\end{center} \end{figure}

This problem is usually solved by setting up $N$ linear
equations between the voltages at nodes (the Kirchoff equations), and
solving them. We only note here the final result: Represent the network as
a graph $G$  with the link between node $i$ and  node $j$ having the
weight  $\sigma_{ij}$. We construct  spanning trees on this graph, and
define the weight of a tree as the product of $\sigma$'s of the occupied
links. We define $T(G)$ as the sum of weights of all spanning trees on the
graph $G$. Clearly $T(G)$ is a homogeneous polynomial of
$\sigma_{ij}$'s of degree $N-1$. Then
\begin{equation}
\label{kirk}
R_{ij} = T(G')/T(G)
\end{equation}
where $R_{ij}$ is the effective resistance between the nodes $i$ and $j$,
and $G'$ is a graph obtained from $G$ by identifying the nodes $i$ and $j$
\cite{resistors}. As a simple check, one can verify that for simple series
and parallel connection of resistors, this formula gives the right answer
[Fig. 9].  That $R_{ij}$ is expressible as a ratio of two determinants may
have been expected, as it is a solution to a linear set of equations. In
the Kirchoff formula [Eq. (\ref{kirk})], each of the determinants is expressed as
a sum over weighs over over all spanning trees of certain graphs.

The two-point correlation function $G_{ij}$ has a simple interpretation in
the language of resistor networks. We consider a resistor network
corresponding to the graph of the ASM, where each bond is a unit resistor.
The sink site is always kept at potential $0$ by definition. Then, the
two-point correlation function $G_{ij}$ satisfies Eq.(\ref{G}), and can be
interpreted as the potential produced at node $j$ when a unit current
fed in at $i$, and the current is taken out from the sink node.

\subsection{Relation to the Potts Model}

The spanning tree problem is known to be a special case of a well-known
model in equilibrium statistical
mechanics: The $q$-state Potts model in the limit
$q\rightarrow0^+$. This model is defined on a graph $G$, with `spin'
variables $\sigma_i$ defined at its nodes taking integer values
between 1 and $q$. The Hamiltonian is
\begin{equation} H[\sigma]=-\sum_{i<j} J_{ij}
\delta(\sigma_i,\sigma_j)\ ,
\end{equation}
where the sum is over the nodes of the graph.
 The coupling $J_{ij}$ is nonzero if and only if nodes $i$ and $j$ are
linked. We write
\begin{equation}
exp[ J_{ij} \delta(\sigma_i,\sigma_j)] =
[1+v_{ij}\delta(\sigma_i,\sigma_j)]\,
\end{equation}
where $v_{ij}=\exp(J_{ij}/T)-1$.
The partition function of the Potts model can then be written as
\begin{equation}\label{partition} Z(q,T)=\sum_{[\sigma]}\prod_{E(G)}
[1+v_{ij}\delta(\sigma_i,\sigma_j)]\ , \end{equation}
where the product is over the set of edges $E(G)$
and the summation takes each spin
variable over its $q$ possible values.

We expand the right-hand-side as a sum of products of the $v$'s. Each term
of this sum can be represented as a graph, in which we represent the term
$v_{ij}$ by the edge $(ij)$ being occupied, and $1$ by the edge remaining
unoccupied. The different terms of the expansion correspond to all
possible ($2^E$, where $E$ is the total number of nonzero $J_{ij}$'s)
configurations of occupied and unoccupied edges of the graph. For any one
such configuration of edges, the sum over $\{\sigma_i\}$ can now be done
explicitly, and we get $q^{c({\bf E'})}$ where $c({\bf E'})$ is the number
of disconnected clusters in the configuration of edges ${\bf E'}$.

Thus the expression (\ref{partition}) can be rewritten in the form, as was
done first by Fortuin and Kasteleyn \cite{kasteleyn}, as sum over the
subsets ${\bf E'}$ of the set of edges ${\bf E}$

\begin{equation} 
Z(q,T)=\sum_{{\bf E'}\subset {\bf E}} q^{c({\bf E'})}v^{{\bf E'}}\ ,
\label{fk}
\end{equation}
where $v^{\bf E'}$ is the product of the weights $v_{ij}$ over all the
edges in ${\bf E'}$.  Being a polynomial in $q$, Eq. (\ref{fk}) may be
used to define the partition function of the Potts model for an arbitrary
non-integer value of $q$. In particular taking the limit $q\rightarrow0^+$
picks out the terms in which there is only one component. We then also
take the limit of small $v$, {\it i.e.}, in the high-temperature limit, we
put $v_{ij}=\beta w_{ij}$, and let $\beta $ tend to zero. Then, this picks
out one component subgraphs with a minimal number of edges, which are just
the spanning trees. In summary, we have the following,
\begin{equation} \lim_{\beta \rightarrow0}
{\beta}^{-N +1}\lim_{q\rightarrow0}q^{-1}
Z(q,T)=T(G)\ . \end{equation}

Here $T(G)$ is the sum of weights of all the spanning trees on the graph
$G$. This equals the number of spanning trees on the same graph, if we put
$w_{ij}=1$ if an edge $(ij)$ is present in $G$, and $w_{ij}=0$ otherwise.
The equivalence to the Potts model is very useful, as a large number of
results are known for the Potts model. An overview of these may be found
in the review by Wu \cite{resistors}.

In two dimensions, the conformal invariance of the critical state
restricts strongly the possible critical exponents, and in fact the
critical exponents of the Potts model in two dimensions are known for all
$q$. For $q \rightarrow 0$ Potts model, the corresponding field theory has
central charge $c=-2$ \cite{conformal}. In this case, most of the
exponents are simple (for example, we have already seen that the
$2$-point correlation function is just the inverse laplacian). There is
one non-trivial exponent which refers to the fractal dimension of chemical
paths in the spanning tree representation. If two points are seperated by
a Euclidean distance $r$, the conformal field theory predicts that the
average length of the path along the tree varies as $r^{5/4}$. Since the
spanning tree describes the propagation of `fire'(i.e. activity) in the
system, the chemical path measures the time taken for the activity to
travel to distance $r$. Thus, we conclude that average time for avalanche
to spread to distance $r$ varies as $r^{5/4}$ in two dimensions.

\subsection{Waves of Toppling}

We have already seen that topplings in an ASM can be done in any order. One
very useful way to relax is by a succession of waves of topplings. Let the
site where new grain is added be $O$. If after addition, $O$ is still
stable, the relaxation process is over. If it is unstable, we relax it as
follows: topple $O$ once, and then allow the avalanche to proceed by
relaxing any unstable sites, without however toppling $O$ again. This
constitutes the first wave of toppling. If at the end, site $O$ is still
unstable, we allow it to topple once more, and let the other sites relax,
until all sites other than $O$ are stable. This is the second wave of
toppling. Repeat as needed. Eventually, site $O$ is no longer unstable at
the end of a wave, and the relaxation process stops.  It is easy to see that
in any wave, the set of toppled sites forms a connected cluster with no
voids (untoppled sites fully surrounded by toppled sites), and no site
topples more than once in one wave. ( This would not be true if the graph
had greedy sites.)

The usefulness of this decomposition of an avalanche into waves of toppling
comes from the fact that the set of all waves is in one to one
correspondence with the set of all two-spanning trees [these are two
disconnected trees that together span the whole lattice] \cite{ikp94}. In
the Potts model formulation these will be the $O(q^2)$ terms in the
partition function in Eq. (\ref{fk}). It is sometimes convenient to work in
the ensemble in which all waves are given equal weight in determining
avalanche statistics. In such an ensemble, we view the evolution of system
as a sequence of waves.  Given a long time series of waves, say from a
simulation, we may ask ``what is the probability that in a randomly picked
wave in this series, there were $s$ topplings?'' This probability will be
denoted by ${\sf Prob}_{w}(s)$. Note that in this ensemble, larger avalanches
having more waves get more weight. 

In two dimensions, it is easily seen that ${\sf Prob}_{w}(s)$ varies as $1/s$
for $1 << s << L^2$, where $L$ is the size of the system\cite{ikp94}. Dhar
and Manna showed that the distribution of the number of sites toppled in
the last wave can be related to number of sites disconnected if a bond is
deleted at random from a random spanning tree \cite{lastwave} . The latter
is expressible in terms of the chemical exponent for spanning trees. Using
the known value $5/4$ of this in two dimensions, they found that the
probability that the last wave in the avalanche has exactly $s$ topplings
varies as $s^{-11/8}$ for large $s$.

\subsection{Random Walks and Loop-Erased Random Walks}

There is a  well-known connection between the resistor-network and
spanning-trees problems and simple random walks. This comes from the fact
that effective resistance between any two nodes in a resistor network can
be expressed in terms of the generating function of simple random walks on
the nodes of that graph \cite{spitzer}.

Consider a random walk on the vertices of a graph $G$ of $N$ vertices. We
imagine the walk to have been going for a long time so that all vertices
have been visited at least once by the walker. At any time $t$, let
$e_j(t)$ be the edge corresponding to the direction of last exit of the
walker from $j$, for all $j$ except the position of walker at $t$. Then
clearly, the set of edges $\{e_j(t)\}$ forms a spanning tree on $G$.

As the walker does a simple unbiassed random walk on $G$, the spanning
tree formed by last-exit bonds undergoes a Markovian evolution in time. It
was shown by Broder \cite{broder} that in the steady state of this Markov
process, all spanning trees are generated with equal probability. [The
argument is quite straight-forward: it is easy to see that in this
Markov process, the transition rate for tree $T \rightarrow T'$ is same
as the rate for $ T' \rightarrow T$. Hence, by detailed balance, in
steady state all trees occur with equal probability.] We thus
have a simple way to generate a uniform distribution of spanning trees on
any graph by using simple random walks. Note that using the one-to-one
correspondence between the recurrent configurations of the ASM and
spanning trees, a time-series of recurrent configurations of ASM obtained
by random grain addition and relaxation also gives a (different) 
time series of spanning trees with uniform distribution.

  A loop-erased random walk is defined as the path of a random walker in which any
loop is erased as soon as it is formed.  Clearly a loop-erased walk has no
self-intersections. As such, the loop-erased walk problem is a more mathematically
tractable variation of the better-known self-avoiding walk problem. It is easy to see
that in a spanning tree formed by last exit bonds, the unique path along the tree from
the origin to end point of the walk is the same as the loop-erased random walk at that
time.

 As loop-erased random walks are the same as of chemical paths on spanning trees,
their geometrical properties are the same. In particular, the fractal dimension of
loop-erased random walks is $5/4$ \cite{lesnm}. One can use this equivalence to relate
other properties of loop-erased random walks to those of spanning trees, hence also to
ASM's \cite{lerw}.  For example, the probability that the next step of the random
walker will result in an erasure of a loop with area $s$ in the sandpile language is
related to the probability that the the last wave will have exactly $s$ topplings. The
probability of forming a loop of area $1$ in two dimensions is also related to the
concentration $f_1$ of sites of height $1$. Let me quote just one more result: If
${\sf Prob}_{\rm lerw}(\ell)$ is the probability that the next step of a loop-erased random walk
forms a loop of perimeter $\ell$, and ${\sf Prob}_{\rm st}(\ell)$ is the probability that adding
a link at random to a randomly chosen spanning tree will form a loop of perimeter
$\ell$, then 
\beq 
{\sf Prob}_{\rm lerw}(\ell) = {\sf Prob}_{\rm st}(\ell)/\ell. 
\eeq
  We shall not discuss loop-erased random walks further here. Interested reader should
consult  \cite{lawler} for further discussion.

\section{Directed Abelian sandpile model}

The abelian group structure of the original BTW model still does not allow
us to deal effectively with the avalanches in the model.  The reason is
that the avalanche statistics is very `coordinate-dependent'. For an
abelian group generated by two operators $\{ a_1, a_2\}$, an equally good
choice of generators is $\{a_1 a_2, a_2\}$. But this set will clearly have
a different avalanche statistics!
 
I describe below a variant of the BTW model, the directed ASM. The model
was defined to take into account the fact that under gravity, particles
would only fall  `down', and not `up' \cite{dr89}; and the existence of
preferred direction usually changes the universality class of critical
behavior. It  has the added advantage that it is mathematically
simpler, and allows  explicit calculation of various quantities.

The model is defined on a square lattice oriented in the $(1,1)$
direction, so that bonds are at $45^\circ$ to the edge. Sand grains are
added anywhere on the top edge with equal probability. For simplicity, we
assume periodic boundary conditions in the horizontal direction (see fig.~10).
On each toppling, one grain of sand is transferred to each of the
two downward neighbors. Particles can leave the system from the bottom.

\begin{figure}
\begin{center} 
\epsfbox{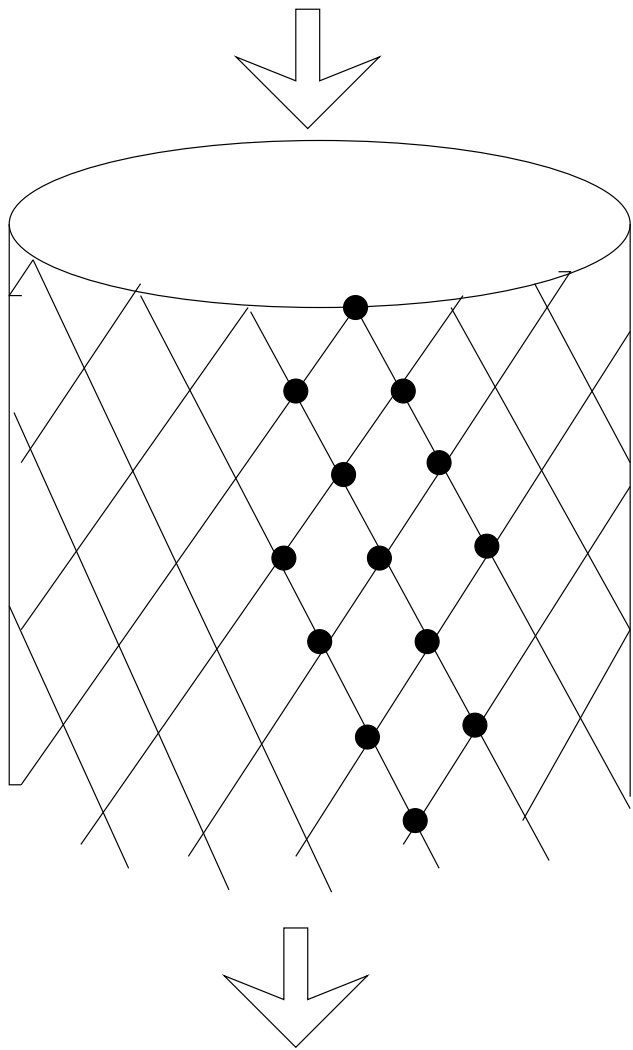}
\label{directed}
\end{center}
\caption{The lattice structure of the directed ASM. Sand is added and
toppled from  top to bottom. Periodic boundary conditiond are used in the
horizontal direction. Note that avalanche clusters have no holes. }
\end{figure} 

  This model can be described as a `child playing with wooden blocks on a
staircase' \cite{kathmandu}. Sitting on the top stair, the child drops the
blocks randomly, and when a block lands on another block, both blocks fall
to the two adjacent sites at the next step below. This perhaps better
models the dynamics of a real sandpile, where most of the sand  stays
inert inside the pile ( playing the role of the staircase), the avalanches
only reorganize the surface layer, and the sand falls  only 
downwards  in topplings.

The toppling matrix $\Delta$ of a directed sandpile is upper triangular.
Thus, $\det \Delta=|{\bf R}|$ is just the product of the diagonal
elements, which is equal to $2^N$. Hence all stable configurations are
recurrent, and the invariant measure is just the product measure of single
site distributions, each one having equal probability of having one grain
or no grains.

This observation permits a direct calculation of the probability of
avalanche sizes. For example, the probability that upon addition of a
particle to an arbitrary site at the top of the sandpile nothing
happens, is just the probability that the site in question has height
0, namely $1/2$. Thus, we get
\begin{equation} {\sf Prob}(s=0)={1\over2}\ .\end{equation}
The probability to get a single toppling is the same as the
probability of having 1 grain in the top site, and no grains in the
two nearest neighbors below as in the following 
diagram,~\unitlength=.1mm\begin{picture}(60,60)\put(30,50){\circle*{20}}
\put(10,0){\circle{20}}\put(50,0){\circle{20}}
\end{picture}. Here, the filled and open circles denote occupied and
empty
sites respectively. Hence,
\begin{equation} {\sf Prob}(s=1)={1\over2^3}={1\over8}\ .\end{equation}
It is straightforward to continue in this procedure. For $s=2$ avalanches two
similar configuration have to be taken into account
\begin{center}\begin{picture}(100,100)\put(50,100){\circle*{20}}
\put(30,50){\circle*{20}}\put(70,50){\circle{20}}
\put(10,0){\circle{20}}\put(50,0){\circle{20}}\end{picture}\qquad
\begin{picture}(100,100)\put(50,100){\circle*{20}}
\put(30,50){\circle{20}}\put(70,50){\circle*{20}}
\put(50,0){\circle{20}}\put(90,0){\circle{20}}\end{picture}\qquad,
\end{center}
so
\begin{equation} {\sf Prob}(s=2)={2\over2^5}={1\over16}\ .\end{equation}

This procedure can be continued for determine ${\sf Prob}(s)$ for any
finite $s$. However, explicit enumeration is tedius if $s$ is not very
small. We now consider the asymptotics of avalanche size distribution for
large values of $s$. To this end we note that if two neighboring sites in
the same level topple, then the site which is directly below  both of
these sites get two sand grains, and must also topple. This implies that
avalanches contain no holes. Therefore an avalanche is completely
determined by its left and right boundaries.


Consider the motion of the right boundary as the avalanche progresses
downwards. Whether the next step is to the right or to the left, is
determined only by the height of the site immediately below and to the
right. If the height in question is one, then step is to the right, and if
it is zero, it is to the left. Thus, the right boundary, and the similarly
the left boundary, are described by simple random walks. The probability
that the avalanche stops at the layer $T$-th layer is the same as the
probability that two random walks in one dimension, which start at the
same point, meet for the first time after a time $T$. This is a classical
problem in probability, and the probability distribution is well-known
\cite{feller}. \begin{equation} {\sf Prob}(\mbox{Avalache stops at the
$T$-th layer)} =\left({\frac{1}{2T+1}}\right)^{2T+2}C_{T+1} 4^{-T-1}\ . 
\end{equation}
For large $T$ this varies as $T^{-3/2}$.

As an object delimited by random walks, the typical width of an avalanche
of duration $T$ scales like $T^{1/2}$, and the typical number of toppled
sites scales like $T^{3/2}$. Therefore, \begin{equation}{\sf
Prob}(\mbox{Avalanche size }>s)\sim {\sf Prob}(\mbox{Duration
}>s^{2/3})\sim s^{-1/3}\ .\end{equation} Finally the probability of
getting avalanche of size $s$ scales like the derivative of the previous
expression, namely as $s^{-4/3}$.

The directed ASM can be solved exactly in higher dimension as well. For $d
>2$, the boundaries of clusters are not lines, but surfaces. Also, the
clusters can have holes. So, we cannot use the simple random walk
arguments to determine the exponents. The tools needed to analyze
avalanche statistics in this general case are the two- and three-point
Green's functions. The simplifying feature of the directed ASM's is that
these functions satisfy linear equations. The two point Green's function
$G_2(t,\vec x|0,\vec 0)$ has already been discussed before: It is the
probability of a topple at the site $(t,\vec x)$ given an addition of
grain of sand at $(0,\vec 0)$. Here we denote by `time' the coordinate
along the down direction, and $\vec x$ is a $(d-1)$-dimensional vector
giving the position of a site on the constant-$t$ surface.

In the invariant state it is equally likely to have any number between
$0$ and $d-1$ grains of sand at a given site, and therefore the
probability that it topples  given that $r$ of its upward neighbors have
toppled is $r/d$. This yields the relation
\begin{equation} G_2(t,\vec x|0,\vec 0)={\left<r\right>\over d}=
{1\over d}\sum_i G_2(t-1,\vec x-\vec e_i|0,\vec 0)\ ,\end{equation}
which is a discrete diffusion equation for $G_2$. Its solution behaves
for large $t$ as
\begin{equation}\label{g2} G_2(t,\vec x|0,\vec 0)\sim {1\over t^{(d-1)/2}}
\exp(-{\vec x^2\over {2t}})\ .\end{equation}
The flux of particles through a surface of equal $t$ scales like $t^0$
as expected
\begin{equation}\mbox{Flux }\sim\int G_2(t,\vec x|0,\vec 0) d^{d-1}x\sim
{\rm const}
\label{flux}\end{equation}

Let us assume that the probability to have an avalanche of `duration'
greater than $t$
scales asymptotically as $t^{-\alpha}$. The flux relation implies that an
avalanche that reaches $t$ has typically $t^\alpha$ topplings there, and
therefore the total mass of the avalanche $s$ varies as $t^{1+\alpha}$. It
follows that the probability of having an avalanche of size greater
than $s$ scales
like $s^{-{\alpha\over1+\alpha}}$.

To find $\alpha$ we need another independent measure, for example
\begin{equation} \left<\left(\mbox{Flux through $t$}\right)^2\right>\sim
t^{\alpha}\ .\end{equation}
The expectation of the square flux is expressible in term of the
$3$-point function $G_3(t,\vec x,\vec y|0,\vec 0)$, which is the
probability of that both $(t,\vec x)$ and $(t,\vec y)$ topple given
that one particle was added at $(0,\vec 0)$. $G_3$ obeys an equation similar to
that of $G_2$,
\begin{eqnarray}\label{g3} 
G_3(t,\vec x,\vec y|0,\vec 0)&=&{1\over d^2}\sum_{ij}
G_3(t-1,\vec x-\vec e_i,\vec y-\vec e_j|0,\vec 0),\mbox{ for~  $\vec x
\ne \vec y$}\\ \label{g3xx}
& = &G_2(t,\vec x|0,\vec 0),\mbox{ for $\vec x = \vec y$.} 
\end{eqnarray}
Eq. (\ref{g3}) is solved by substituting the ansatz
\begin{equation} G_3(t,\vec x,\vec y|0,\vec 0)=\sum_{t'=0}^t\sum_{\vec z}
f(t',\vec z)G_2(t,\vec x|t',\vec z)
G_2(t,\vec y|t',\vec z)\ ,\end{equation}
where the sum is over all the sites that may be toppled by an
avalanche starting from $(0,\vec 0)$. The unknown function $f(t,\vec z)$ is
constructed recursively, so as to satisfy the boundary conditions
eq.~(\ref{g3xx}).
Putting $\vec x = \vec y$ in equation~(\ref{g3}), we get
\begin{equation} 
\label{detf}G_2(t,\vec x|0,\vec 0)=\sum_{t'=0}^t\sum_{\vec z} f(t',\vec
z)G_2^2(t,\vec
x|t',\vec z)\ , 
\end{equation}

This equation expresses the known function $G_2(t,\vec x|0,\vec 0)$ as a sum of
terms linear in $f(t',\vec z)$ where the site $(t',\vec z)$ is in the
backward light cone of $(t,\vec x)$. Hence these can be solved recursively
for all $ f(t,\vec z)$, starting with $t=0$.

The flux condition~(\ref{flux}) plus the
boundary condition~(\ref{g3xx}) yield a sum rule 
\begin{eqnarray} 1/d&=&\sum_{\vec x}G_2(t,\vec x|0,\vec 0)=
\sum_{\vec x,\vec z}\sum_{t'=0}^t f(t',\vec z)G_2(t,\vec
x|t',\vec z)^2 \\
 &=& \sum_{\vec z}\sum_{t'=0}^t
f(t',\vec z)K(t-t')=
\sum_{t'=0}^t g(t')K(t-t')\ , \end{eqnarray}
where we have defined $g(t)=\sum_{\vec x}f(t,\vec x)$, and 
$K(t)=\sum_{\vec x}G_2^2 (t,\vec x|0)$. The kernel $K$ is easily
calculated in terms of the known $G_2$ [Eq.~(\ref{g2})], and it scales for
large $t$ as
\begin{equation} K(t)\sim t^{-{d-1\over2}}\ .\end{equation}

For $d > 3$, $\sum_t K(t)$ converges, which shows, using the sum rule
that $g(t)$ tends to a constant for $t$ large. When $d=3$ 
$\sum_t K(t)$ diverges logarithmically so $g(t)\sim 1/\log t$ and,
similarly, in two dimensions $g(t)\sim t^{-1/2}$. Using these results
and
\begin{equation} \left<\left(\mbox{Flux through $t$}\right)^2\right>=
\sum_{t'=0}^t
g(t')\end{equation}
then shows that $\alpha = 1$ in dimension $d > 3$. In $d=3$, we get
$g(t) \sim 1/(log t)$. This implies that the probability that avalanche
stops at the $T$-th layer varies as $ (log T)/T^2$. These logarithmic
corrections have been verified in simulations \cite{logcorrect}.
We have already seen that in two
dimensions, $\alpha=1/2$. The upper critical dimension is three.

\section{Relation to Other Models}
In the previous section, we  studied the DASM as an example of a
self-organized system
in which the invariant state is easily
charcterized. Another motivation for studying this model is that it
equivalent to other statistical models which were proposed in other
contexts. Here I will discuss three such models.

\subsection{Scheidegger's Model of River Basins}

The first is the Scheidegger's model of river basins \cite{rivers}.
Scheidegger, a hydrologist, originally proposed this model as a very
simplified description of how rainfall accumulates into small streams
which then coalesce to become larger and larger river.  We consider a
large area with roughly uniform average rainfall, and having a constant
average slope (say southward).  We represent the area approximately as a
square grid, tilted and oriented as in the DASM. At any point of the
lattice, there may be zero, one or two streams of water bringing water
from its two upstream neighbors. If $n$ units of water come into a
site per unit time (usually taken to be a year in geophysics), $(n+1)$
units of water leave the site in a single stream towards one of the two
sites just below it (the extra one unit comes from the local rainfall).
Whether it goes to the south-east or south-west neighbor depends on local
geographical details. In this model, these directions are chosen randomly
at each site.

If we draw the directions of water-flow going out of each site, we get a
picture of the river network of the area. Such networks are also called
drainage networks. [See Fig. 11]. In a drainage network, there is a
unique path from each site to the `sea' (the lower boundary in Fig. 11).
Also, there are no loops, if we ignore the possibility of a single stream
breaking into two streams (fairly good approximation away from delta
regions), the drainage networks must form a directed spanning tree.

\begin{figure}
\begin{center}
\leavevmode
\psfig{figure=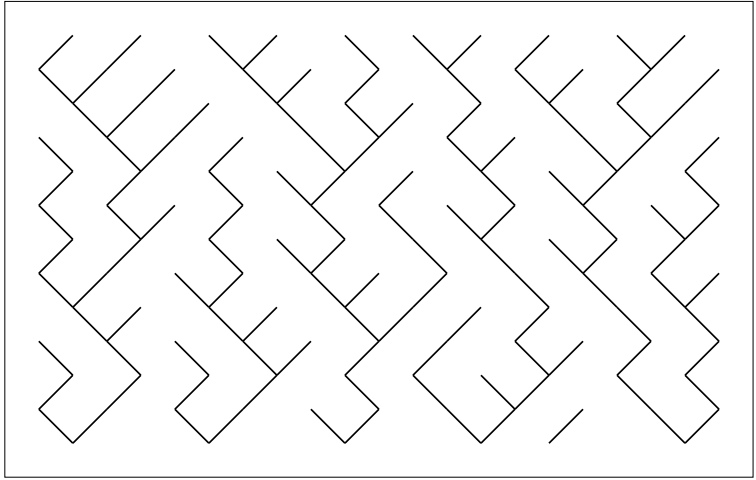,width=10cm}
\caption{ The Scheidegger model of a river network.}
\end{center}
\end{figure}

Using these rules it is now possible to calculate the probability that the
outflow from a site is one unit: This event happens if and only if the
site receives no water from either of its two up neighbors, because the
right hand side neighbor has outflow directed to the right, and the left
neighbor is flowing to the left. Thus ${\sf Prob}({\rm outflow}=1)=1/4$~.
Similarly, to get an outflow of two units, it is necessary to have inflow
of one unit, so that one of neighbors from above has to have outflow of 1
directed at the given site, and the other neighbor from above has to have
outflow directed to its other neighbor from below. Such an event has
probability $1/16$, and since there are two ways to choose the the flow
into the site, we have ${\sf Prob}({\rm outflow }=2)=1/8$.

On comparing the previous argument with the discussion of the DASM we see
that the configuration of sites required to get an outflow of two units is
identical to that which is associated with an avalanche of size two,
starting from the given site and going {\it upwards}. The two
configurations which are associated with outflow of two units are the ones
possible for a size two avalanche, and a similar result holds for larger
values of the outflow. The only difference is that there is no possibility
of outflow of zero units. Therefore we have the relation \begin{equation}
{\sf Prob}({\rm outflow }=s)=2\,{\sf Prob}(\mbox{DASM avalanche size
}=s),\quad \mbox{for $s>0$.}\end{equation}

Now all the earlier discussed results regarding the DASM apply equally
well to the river network model. In particular, a river of total length
$\ell$ will drain typically an area of size $\sim\ell^{3/2}$. This
observation is in rough agreement with observation on rivers on scales
which are not very large.

We can study the model in higher dimensions. In three dimensions, one can
think of the capillary network for flow of blood in an animal, where each
`cell' generates `waste', which has to be transported out using the
tree-like network of capillaries. Drainage networks in higher dimensions
are harder to visualize.

The Scheidegger river network model in fact generates a directed spanning
tree of the lattice. When the river network is situated in a relatively
flat landscape, it should resemble more an undirected spanning tree, so a
simple model for such a network would be just to assume that in a river
network on a flat landscape, all spanning trees are equally likely to be
realized. {\it The geometrical structure of river networks thus can be
modelled by that of a random spanning tree in two dimensions.} It turns
out that this simple model describes the real river networks just as well
as the more complicated models proposed in literature.

The random spanning trees are described by the $0$-state Potts model.
Using the results from the conformal field theory for Potts model, we see
that for random spanning trees, the typical length $L$ of the path along
tree between two points at a Euclidean distance $R$ scales like $L\sim
R^{5/4}$.  This scaling provides an estimate of the area $A$ drained by a
river of length $L$ \begin{equation} A\sim R^2\sim L^{8/5}\ ,
\end{equation} that is, slightly larger than the exponent in the directed
network (where $A \sim L^{3/2}$).

There is a large amount of literature dealing with the modelling of river
networks. If we include the phenomena of erosion, which transports matter,
and thus alters the elevation across the landscape, which in turn
affects the river-flow, we have a complex self-organized structure. A long
excursion into models of river networks is outside the scope of these
lectures. I will only provide a few pointers here \cite{refrivers}.

\subsection{ Takayasu's Model of Diffusion with Aggregation}

Another model, which models a physically very different phenomenon, but
falls into the same equivalence class is the Takayasu aggregation model
\cite{takayasu}. It describes a system in which particles are
continuously injected , diffuse, and coalesce. In its simplest example, it
can be
defined on a one-dimensional chain. The explicit rules of evolution are:
\begin{enumerate}
\item At each time step, each  particle in the system  moves by a single
step, to the left or to the right, 
taken  with equal probability, independent of the choice at other sites.
\item A  single particle is added at every site.
\item If there are more than one particles at one any site, they coalesce
and become a single particle whose mass is the sum of the masses of the
coalescing parts. In all subsequent evolution, the composite particle acts
as a single particle.
\end{enumerate}

A quantity of interest in this system is the
probability distribution of the total mass at a randomly chosen site at
late times. Since mass is added all the time and cannot escape, the
average mass per site is proportional to $t$. As discussed earlier, a diverging first
moment of a distribution could come from a  power law tail, with a cutoff which increases as a
power of the time $t$. Such a tail indeed exists in the mass distribution.

It is not hard to see that this model in one dimension is very similar to
the (two-dimensional) river network model. We consider a space-time
history of the evolution. The world-lines of a diffusing particles look
like the rivers of Fig 9.  The probability to have a particle a mass $M$
is therefore equal to the probability to have outflow $M$, which is twice
the probability of finding an avalanche of size $M$ in the respective
DASM. Therefore, for large $t$, the mass probability distribution is
characterized by the same exponent

\begin{equation} 
{\sf Prob}(M)\sim M^{-4/3}f\left({M\over M^*(t)}\right)\ ,
\end{equation}
where $M^*(t)$ varies as $t^{3/2}$, and $f$ is some cutoff function. This model is
also easily generalized to higher dimensions.

The river network and aggregation models differ from the sandpile models
in that the driving is finite and deterministic and the evolution is
random. They thus provide examples of systems which display SOC without
the requirement of infinitesimal driving. Sometimes, it is argued that
infinitesimal driving should be considered as a defining characteristics
of SOC \cite{grinstein}. If this is done, Takayasu aggregation model,
while mathematically equivqlent to the sandpile model, which is the
prototypical model of SOC, would not show SOC!

\subsection{ The Voter Model}

The voter model is defined as follows: We consider a $d$-dimensional
lattice. At each site, there is an individual (voter) who has an opinion
on an issue. We take this opinion to be a two-valued discrete variable
(yes or no). A voter can communicate with his neighbors, and can change
his opinion in time. We take time-evolution to be discrete time Markovian:
If at time $t$, a site has $r$ neighbors who vote yes, at time $t+1$, this
site would choose to vote `yes' with probability $r/(2d)$, and `no' with
probability $1 - r/(2d)$. 

Given the particular choice of local transition rates, one can try to
write down evolution equations for ensemble averaged quantities, for
example $\left< \eta(x,t)\right>$, where $\eta(x,t)$ specifies state of
the voter at site $x$ at time $t$, taking value $1$ if `yes', and $0$
if `no'. Normally, such evolution equations involve higher-spin
correlation functions. This gives the well-known BBGKY hierarchy of
equations for correlation functions. It is easy to see that for the 
voter model, this hierarchy closes on itself. The equation for evolution
of   $\left< \eta(x,t)\right>$ involves only the  $\left<
\eta(x',t)\right>$ at neighboring sites $x'$, and no higher order
correlation functions. Specifically, in two dimensions
\begin{equation}
\left< \eta(x,t+1)\right> = {1 \over 4}\sum_{x'} \left< \eta(x',t)\right>
\end{equation}
where the sum over $x'$ is over the nearest neighbors of $x$. These are
linear equations, easily solved by eigenmode decomposition. One can
similarly show that equations satisfied by the 2-point correlation
functions also close on themselves, and so on. This model has been studied
a lot, and a fair amount of literature exists about it \cite{votermodel}.

To note the connection with the directed ASM, consider evolution of a
$d$-dimensional voter model which at time $t=0$ has only one voter having a
`yes' vote in a sea of `no's. Then in the space-time history of the
evolution, there is a single cluster of `yes' sites. It is easy to see
that the probability that this cluster will have a given shape is exactly
the same as that of the same avalanche cluster in the DASM in $d+1$
dimensions.

\section{Solvable Models of Undirected Sandpiles}
\subsection{One Dimensional Chains and Ladders}

The simplest undirected model is defined on a one dimensional chain of
length $L$. The matrix $\Delta$ is tridiagonal
\begin{equation}\Delta=\left(\begin{array}{cccc}2&-1& & \\
-1&2&-1&\\&-1&2&\ddots\\&&\ddots&\ddots\end{array}\right)\end{equation}
and its determinant is easily seen to be  
\begin{equation}\det\Delta = L+1
\ .\end{equation}

It is thus a degenerate case where the number of recurrent configurations
grows only polynomially and not exponentially with $L$. It is not hard
to identify the $L+1$ recurrent configurations. They are
either a configuration with all ones, or a configuration with a single
zero somewhere along the chain.

\begin{figure}
\begin{center}
\epsfbox{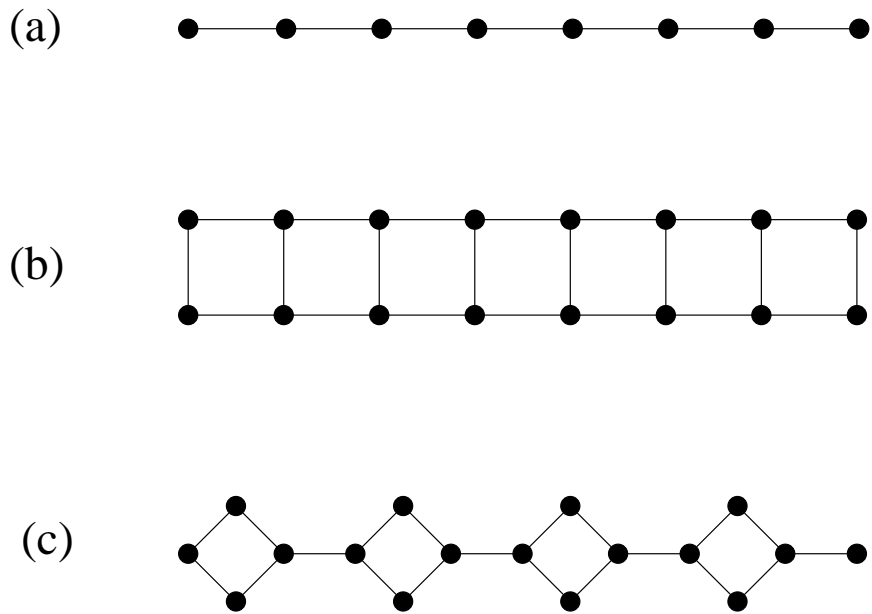}
\label{chains}
\end{center}
\caption{Some one-dimensional lattices: (a) a simple chain (b) a ladder
(c) a necklace.}
\end{figure}

The result of an addition of a grain of sand at point $y$ to a
configuration with a zero at $x$ is as follows: Let us consider first the
case $ x< y$. The site $y$ is
toppled, then its nearest neighbors, and so on, creating a wave which
stops on one side at $x$ and at the other side at $L$. The result of
this wave of toppling is a configuration with two grains of sand at
$y$, zero at $x+1$ and at $L$, and one grain everywhere else. Now a
second wave of toppling begins at $y$ at the end of which the zeros
move to $x+2$ and $L-1$. This process continues until one of the zeros
reaches $y$ and then it stops. The total number of topplings $s$ in such
an avalanche is $s(x,y)= (L-x)\min(L-y,y-x)$. The case $ x> y $ is
similar.  We then obtain 
the avalanche distribution using the fact $x$ and $y$ are uniformly
distributed.
\begin{equation}
{\sf Prob}(s) = \sum_{x=0}^L \sum_{y=1}^L \frac{1}{L(L+1)} \delta(s-s(x,y))
\end{equation}
Details may be found in the paper by Ruelle and Sen \cite{ruellesen}. It
may be of interest to note that for finite $L$, the function ${\sf
Prob}(s)$ is not a monotonically decreasing function of $s$, and shows
gaps  (depending on $L$) where ${\sf Prob}(s)$ is exactly zero.
For large $L$, $x$ and $y$ are of order $L$, and hence $s=O(L^2)$. Another
way to express this is by a direct calculation of the probability
distribution of $s$ which takes the scaling form

\begin{equation} {\sf Prob}_L(s)\sim{1\over L^2}f\left({s\over
L^2}\right)\ , 
\end{equation} 
where $f$ is a scaling function whose
explicit form is known \cite{ruellesen}. The gaps in the function
${\sf Prob}_L(s)$ are filled by the smearing, and the scaling function $f$ is
well behaved. This distribution function has the undesirable feature that
in the thermodynaimc limit $L\rightarrow\infty$ the probability of
observing any finite avalanche goes to zero.

This behavior is atypical even for one-dimensional models. For example, on a
ladder, or  a necklace (a decorated chain)  [Fig. 12],
we can determine the  behavior of successive waves of toppling,
and hence the distribution of avalanches \cite{ad95}. Unlike the linear
chain, on a ladder or necklace, the number of recurrent configurations grows
exponentially with $L$. The avalanches are of two types: Type I consist of a
single wave which stops after toppling a finite fraction of sites, and Type
II which consist of many waves of topplings and is similar to the avalanches
in a single chain [see Fig. 13].  Both these types occure with $O(1)$
probability, so that the event size probablity distribution takes the
asymptotic form

\begin{equation} 
\label{lc2ssf}
{\sf Prob}_L(s)\sim{1\over L}f_1\left({s\over L}\right)+
{1\over L^2}f_2\left({s\over L^2}\right)\ . \end{equation}

\begin{figure}
\begin{center}
\leavevmode
\psfig{figure=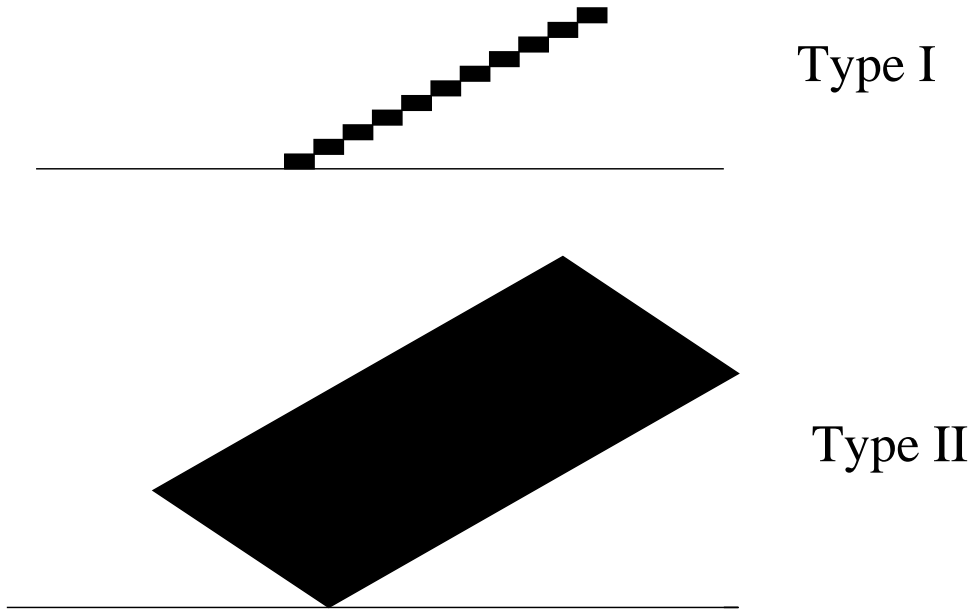,width=8cm,angle=-90}
\end{center}
\caption{The space-time evolution of two different types of avalanches
on decorated linear chains.}
\end{figure}

This linear combination of simple scaling forms is a very simple
example of multifractal behavior, with just two exponents. It
demonstartes how simple finite-size scaling may be violated.

\subsection{ The Bethe Lattice}

Apart from the linear chain and ladders,  the only 
explicitly solvable realization of the undirected sandpile model is
when it is defined on a tree, such as the Bethe lattice. As expected, the model 
displays a  mean field behavior on this
effectively infinite-dimensional lattice. 

I only indicate the general technique of solution. For details, the reader
has to consult the original paper \cite{bethelattice}. Consider the ASM on
a Cayley tree. One starts by characterizing the recurrent configurations.
Consider the subconfiguration on a subtree of $r$ generations. The allowed
configurations can be divided into two classes: weakly allowed, and
strongly allowed. Strongly allowed have no FSC even if it is part of a
bigger tree.  Let $S_r(z)$ and $W_r(z)$ be the number of such
configurations for a subtree with $r$ generations having height $z$ at the
top site of the strong and weak type respectively. It is easy to write
down recursion $S_{r+1}(z)$ and $W_{r+1}(z)$ in terms of their values at
the $r$th generation. For large $r$, these diverge as $\exp(\exp(r))$, but
their ratios tend to finite limits. These limits are used to determine the
relative probabilities of various subconfigurations deep inside the tree
in the steady state. For example, we find that for a 3-coordinated
lattice, the sites with heights $1,2 $ and $3$ occur with relative
frequencies $1:4:7$. The propagation of avalanches is simple in this
lattice. On adding a particle, all sites belonging to the cluster of
sites having height $3$ containing the point of addition topple, and no
other. There are very few multiple topplings. It is then straight forward
to write down explicit expressions for ${\sf Prob}(s)$ for $s=1,2,\ldots$.

 Unlike the one-dimensional
chain, here the probability to get a finite event remains finite in
the thermodynamic limit, the and the probability distribution of the
number of toppling $s$ behaves asymptotically as $s^{-3/2}$. In fact,
the spread of avalanche on the Bethe lattice is very similar to that of
infection in a contact process at criticality,

The mathematical treatment simplifies considerably if we consider
a directed Bethe lattice, such that the number of `in' and `out' edges at
every node is equal to $n$. Then, all stable configurations are recurrent,
and the probability that a site catches infection from an infected
neighbor is precisely $1/n$, same as in the critical contact process
\cite{peng}. From the known results about the critical contact process on
a tree it follows that ${\sf Prob}(s)$ for large $s$ varies as $s^{-3/2}$. Note
that we showed that the ASM which belongs to the $q=0$ Potts model
universality class, but on a Bethe lattice, it becomes equivalent to the
percolation problem, which actually corresponds to $q=1$ Potts model.

There are many other treatments of the mean-field theory of sandpiles,
all give the same set of mean-field exponents \cite{meanfield}.

\section{The Undirected ASM in Two Dimensions}

The undirected ASM on the square lattice is undoubtedly the most studied of the SOC
models. Many results are known exactly, but no closed-form expression for the
avalanche probabilities, or avalanche exponents.

Let us start by summarizing the known results. Consider a $L \times L$
lattice. The number of recurrent states $|{\bf R}|$ is equal to $\det
\Delta$ using Eq. (21). The matrix $\Delta$ is easily diagonalized by a
Fourier transformation, and we get 
\beq 
det \Delta = \prod_{\ell=1}^{L}
\prod_{m=1}^{L} [4-2 cos ({\frac {2 \pi \ell}{L+1}}) -2 cos ({\frac {2 \pi
m}{L+1}})] 
\eeq This number grows as $\exp(\omega  L^2)$ for large
$L$, where $\omega $ may be called the entropy per site in the
steady state. For large $L$, we may replace the product by an exponential
of an integral, giving
 
\begin{equation}
\omega = \int_0^{2 \pi} { d\theta \over {2 \pi}} \int_0^{2
\pi} {d\phi \over {2\pi}} ~log( 4 - 2 cos\theta -2 cos\phi)
\end{equation}
 
\subsection{ Height Correlations in the Steady State}

For this case, a fairly detailed characterization of the steady state has been
achieved. For example, we may wish to find the probability distribution of the height
at a given site, as it determines the probability that there is no toppling if add a
particle. The simplest to calculate is \cite{md91}

\begin{equation} {\sf Prob}(z_i=1)={\#\{\mbox{recurrent states
in which $z_i=1$}\}\over \#{\bf R}}\ .\end{equation}

Consider an auxialiary ASM defined on a graph with site $i$ and its
connecting links removed. In the burning algorithm, a site
with $z_i=1$ can be burned only after all of its neighbors have been
burnt. Let ${\cal C}$ be a recurrent configuration of the sandpile with
$z_i=1$. We define ${\cal C'}$ to be a sandpile configuration of the
auxiliary ASM obtained from ${\cal C}$ by deleting the site $i$, and
decreasing the height at each neighbor of $i$ by $1$. It is easy to see
that burning proceeds identically in ${\cal C}$ and ${\cal C'}$. Thus
there is one-to- one correspondence between recurrent configuration of the
original ASM with $z_i=1$, and all recurrent configurations of the
auxiliary ASM. But the latter is easily calculated by the determinant of
its toppling matrix, say $\Delta'$.

We
augment the toppling matrix $\Delta'$ by adding a diagonal entry $1$ for
the site $i$. The only rows and columns in which $\Delta'$
is different from $\Delta$ are those associated with $i$ and its four
neighbors. Thus we can write $\Delta'=\Delta+\delta$ where $\delta$ has
non-zero entries only in a $5 \times 5$ sub-matrix. Then
\begin{equation}
{\sf Prob}(z_i=1)={\det \Delta'\over\det\Delta}=\det(1+\Delta^{-1}\delta)
\end{equation}
which boils down to the calculation of a $5 \times 5$ determinant.
The outcome of this calculation  \cite{md91} in the limit
$L\rightarrow\infty$ is
\begin{equation}
{\sf Prob}(z_i=1)\equiv f_1={2\over\pi^2}(1-{2\over\pi})\ \simeq 0.0736.
\end{equation}

One can also calculate the probability that in the steady state, a
randomly chosen site has height $2,3$ or $4$. These require a more
sophisticated use of graph-theoretical techniques. As this calculation is
somewhat involved \cite{p94}, (even writing down the the exact analytical
expressions for the final result would take up nearly a page) we only
quote the numerical values of the final result here : The concentration
sites of heights $2, 3$ and $4$ in the steady state are approximately
$0.1739, 0.3063$ and $0.4461$ respectively.

Similarly it is possible to calculate as a ratio of determinants the joint
probability that two sites $i$ and $j$ both have  height $1$. The calculation now
involves
calculating the determinant of a $10 \times 10$ matrix. If the distance $r_{ij}$
between the sites is large, the leading r-dependence is found to be  of
the form

\begin{equation}
{\sf Prob}(z_i=1,\,z_j=1)\sim f_1^2+{A\over r_{ij}^4}\ .
\end{equation}

[More generally, in $d$-dimensions, the correlations in heights decay as $r^{-2d}$
with the seperation $r$.] The coefficient $A$, which may be calculated explicitly, is
negative. Thus, the height variables show some anticorrelation. We recall that having
two adjacent sites with height $1$ is  forbidden in the steady state.

From the clustering property of correlation functions, one  
expects that the general expression for the height-height correlation function,
consistent with the above form, will be of the form
\begin{equation}
{\sf Prob}(z_i=a,\,z_j=b)\sim f_a f_b + {g_a g_b\over r_{ij}^4}\ .
\end{equation}
Summing this probability over $b$ ranging from $1$ to $4$, we must
get $f_a$, the concentration of sites with height $a$. This implies
that we must have $\sum_{a=1}^4 g_a =0$, in addition to the trivial
constraint $\sum_{a=1}^{4} f_a =1 $. The calculation of $g_a$ seems to be
rather complicated, and has been accomplished so far only when site $i$ and
$j$ are the boundary sites \cite{ivashsur}. 

One can also study how the height probabilities are modified near an open
boundary. It is found that mean height is lesser than the bulk value at a
distance $z$ from the boundary, by an amount which varies as $z^{-2}$
\cite{ivashsur}.

 \subsection{ Avalanche Exponents in Two Dimensions}

We now come to the still unresolved question of exponents of avalanches in
the 2-d model. The avalanche  exponents, which are expected to
be universal in two dimensions, are still unknown.  The values of numerical
estimates made by different people have shown a wide range of values
\cite{exponents}, with different methods of analysis giving different
values.  Avalanche sizes may be measured in  terms of different quantities, 
such as  the total number of topplings $s$, the number of
distinct sites toppled $s_d$, the radius $R$ of the affected area, the
`duration' $T$ of the avalanche, and $n_w$, the number of waves in the
avalanche.

Let us start with the  simple finite-size scaling picture (this may have to
modified later). In this picture, these measures are sharply peaked
functions of each other, and given the diameter of avalanche $R$, other
measures scale as simple powers of $R$. In this case, we would expect
that
\begin{equation}
{\sf Prob}(R) \sim r^{-\tau_r}f(R/L), 
\end{equation}
where $L$ is the size of system.    

 From the compactness of avalanche clusters, it follows that $s_d$ scales
like $R^2$. From the Potts model equivalence, we see that the time $T$
needed to reach a site of distance $R$ from the origin of the avalanche
scales as the chemical length of the path, so that $T\sim R^{5/4}$.  If
$\tau_s,\,\tau_d,\, \tau_r$ and $\tau_t$ are the exponents characterizing the
distributions for $s, s_d, r$ and $T$, this implies that

\begin{equation}
\tau_r - 1 = 2 (\tau_d -1) = {5 \over 4}(\tau_t -1)
\end{equation}

Let us assume that $n_w$ scales as a power of $r$, say as $r^y$. Then, $s$
varies as $n_w s_d$, and hence as $r^{2+y}$.  To determine the avalanche
exponents, we use two relations:  $<s> \sim L^2$, and $<n_w> \sim log L$.  
These immediately give us 
\begin{equation}
\tau_r =y +1,\qquad \tau_s  = 2(1+y)/(2+y)
\end{equation}   
Thus all the exponents are determined up to the
single parameter $y$.

Priezzhev et al using some identification of known scaling dimensions of the
2-dimensional Potts model, have  conjectured \cite{pki96,kp98} that
$y=1/2$ which leads to $\tau_d=5/4$ and
$\tau_s=6/5$ (where $\tau_s$ is the power law decay of the $s$
distribution), which is consistent with the numerical results.

The simple scaling assumption has been challenged recently by De Menech et
al \cite{dmst98}. These authors, based on their
analysis of simulations data, have argued that
avalanches that reach the boundary have a
qualitatively
different behavior than those that do not dissipate particles. They argue 
that the fractional number of avalaches whose radius exceeds $r$
decreases as $r^{-1/2}$, for $r > > 1$. The avalanches that do not reach
boundary have $s \sim s_d$ and $n_c \sim 1$. For
a lattice of size $L$, only a fraction $L^{-1/2}$ of the avalanches reach
the boundary, but those that do, typically have $L^{1/2}$ more waves, so
that the average size of avalanches that reach boundary is $L^{5/2}$.
Thus, according to these authors $y=1/4$ for dissipative avalanches, but
$y \approx 0$ for the non-dissipative ones. 

 This behavior is more
in line with the linear combination of scaling forms discussed earlier [
Eq.(\ref{lc2ssf})]. If the simple scaling approach breaks down due to the relatively
rare large events dominating the mean values of $s$ and $n_w$, we cannot use
the exactly known $L$-dependence of $\left<s\right>$ and 
$<\left<n_w\right>>$ to determine the
avalanche exponents.  These then provide only lower bounds to $\tau_s$ and
$\tau_r$  defined for avalanches whose size is not comparable to the size
of the lattice.

We have already mentioned that the probability of last wave of size $s$
varies as $s^{-11/8}$. Another exact result known is the distribution
of size of a wave generated at the corner of a wedge of angle $\theta$, with
particles allowed to leave system at boundaries. In this case, Priezzhev et
al showed that probability of avalanche of size $s$ varies as
$s^{-\tau(\theta)}$, where $\tau(\theta)= 1+ {\pi \over {2\theta}}$
\cite{wedge}. For a boundary avalanche, this gives $\tau(\pi)=3/2$.

 Recently, the scaling picture of avalanches has been extended to study
correlations between successive 
waves of toppling 
\cite{pb97}. On the basis of extensive numerical simulations, these
authors have made the interesting observation that  the conditional
probability ${\sf Prob}(s'|s)$ that the $(k+1)$-th wave is of size
$s'$, given that the previous wave had $s$ topplings is roughly
independent of $k$. This varies as $(s'/s)^a A(s)$ for $s' <<s$, and as
$(s'/s)^{-b} B(s)$ for $s'>>s$. Here $A(s)$ and $B(s)$ are some
$s$-dependent amplitudes. However, the range of validity of these power-laws
is limited to $1 << s, s' << L^2$.

\section{The Abelian Distributed Processors Model}

It is useful to see how far ASM can be generalized, still retaining its
abelian group structure. In this section, I shall discuss an interesting
generalization of the ASM. The models consist of a network defined on a
graph, on whose nodes sit processors, which are finite state automata.
Each of these has an input stack of messages. When the input stack of a
processor is not empty, it pops one of the messages according to a
predetermined order, and processes it. As a result of the processing its
internal state may change, and some messages may be transmitted to its
neighbors in the network, or to the outside world.  Then if the input
stack is still not empty, the next message is processed, otherwise the
processor idles. The processing rules for different processors may differ.

In many applications, especially in computer science, one considers
such networks where the speed of the individual processors is unknown,
and where the final state and outputs generated should not depend on
these speeds. Then it is essential to construct the network so that
the order at which messages arrive to the processors is immaterial.
If this is indeed the case, then the netwrok may be called abelian,
and the model is named an abelian distributed processors (ADP) model.

Evidently, the ASM is an example of an abelian distributed processors
model. Another condition which may be imposed on an ADP is the equivalent
of the `good behavior' conditions for the ASM. Namely, that the response
of the network to a single message sent to any of the processors
terminates in a finite time, after which all the processors sit idle until
further input is applied. The state in which all processors are idling,
and all input stacks are empty is called a quiescent state. The quiescent
states are the analogs of the stable configurations of the ASM.

As in our discussion of the ASM, one may define generators $a_{i,m}$
whose action is to send message $m$ to processor $i$. These generators
commute, and generate an Abelian semi-group, acting on the set of
quiescent states; restricted to the set of recurrent states, they form
an abelian group.

\subsection{ The Eulerian Walkers Model}

As an example of an ADP, consider the following model defined on a square
lattice of finite size $L$ \cite{pddk96}. On each site sits an arrow
pointing toward one of the four directions, up, down, left or right. A
walker is placed on the lattice at one of the sites. The walker reaching a
site resets the direction of the arrow at that site by rotating it by
$90^\circ$ clockwise, and takes a step to the neighboring site in the new
direction. It repeats this action, till it leaves the lattice.

We add a walker at a randomly chosen site, wait till it
leaves
the system, and then add another walker. It is easy to see that this model
is an example of the ADP model.  We can define commuting operators
$\{a_i\}$ whose action on a configuration of arrows results in a new
configuration of arrows obtained by adding a walker at $i$ and allowing
it
to move till it leaves the system. In fact, the algebra of the operators
$a_j$ of introducing a walker at site $j$ is [Fig. 5]
\begin{equation}
a_j^4= a_{j_1} a_{j_2} a_{j_3} a_{j_4}\ , \end{equation}
which is identical with the algebra of the ASM on a square lattice.
Therefore we can apply the variation of the script test to see if a
particular configuration is recurrent. However, it is  easy to see
directly that in a recurrent configuration the arrows form a directed
spanning tree on the lattice.

Although the algebra is identical, the avalanches of this model are
different from those of the ASM. Since the configuration does not
become quiescent before the walker leaves the system, the typical
avalanche size is $O(L^2)$ in contrast with the $O(L^0)$ avalanches
which are most likely in the ASM. Thus we see again that the group
structure
of the algebra of operators cannot by itself determine the avalanche
exponents.

Several quantities of interest in this model have been studied. On an $L
\times M$ torus, such a walker eventually ends up in a limit cycle, whose
period is exactly $ 4 L M$, and it traverses each bond of the lattice
exactly twice in one cycle, once in each direction . Such a tour of the
lattice is called an Eulerian circuit, hence the name of the model.  If a
walker starts at origin in an initially random configuration of arrows, it
reorganizes the arrows as it moves, and in low dimensions retraverses the
already visited sites many times.  Its mean square deviation in time t
increases as $t^{\frac{2}{d+1}}$ for $d \leq 2$ and as $t$ for $ d>2$.
  
To make the connection to the ASM more explicit, consider a model in
which we drop walkers on the board. Each walker waits at the site where
it is, except that if  the number of walkers waiting at a site becomes 
greater than
or equal to  $r$, $r$  walkers leave the site in consecutive
clockwise directions, and the arrow is reset. The case $r=1$ corresponds
to the Eulerian walkers model The case $r=4$ corresponds to the ASM. In
this case the arrow is always reset to the same direction, and 
it does not matter which way. Thus, we may forget about the arrows, and
recover the ASM.
 
In this intermediate case of $r=2$, small avalanches are possible, but the
critical occupation number is two rather than four as in the ASM. All
these models have the same operator algebra. The analytical treatment of
the $r=2$ model is surprizingly difficult , even on a Bethe lattice, and
the full solution has not been possible so far \cite{shcherbakov}.

\subsection{ The Manna Model}

Another variation which falls in the ADP category, is that of
stochastic ASM \cite{manna91}. In this model the critical height is
two, but at each toppling, the
neighbors to which  the two grains are transferred are chosen at random 
with equal probablity to be either the
up-down neighbors or the east-west neighbors. Such dynamics may be
performed by an ADP by placing a pseudo-random number generator at
each site, and using its output to choose the destination of the
ejected sand. However, it is more convenient to think of the operation
of toppling as being really random, so that the action of the opertor
$a_i$ of adding
a grain of sand at site $i$ does not yield a unique configuration, but
rather a probablity distribution of configuration. Using the vector
space represention defined above for the ASM with random driving, we
see that the algebra of $a$'s becomes [Fig. 5]
\begin{equation}
\label{sasm} 
a_j^2={1\over2}( a_{j_1}a_{j_3} + a_{j_2}a_{j_4}) 
\end{equation}
Thus the operator algebra  gains
richer structure and it becomes a commutative ring.

The set of equations~(\ref{sasm}) written out for all the sites of the
lattice yields a system of quadratic equations in many variables. This
system is much more difficult to analyze than the system of linear 
equation for the deterministic ASM, and not much is known about its
solutions.

\section{Open problems}

Let me conclude by discussing some of the interesting open problems in
this field. The list is necessarily incomplete, and influenced by personal
taste and prejudices.

We can start with the problem of directed abelian sandpile models. In this
case, we can calculate all the avalanche exponents in all dimensions.
However, time-dependent correlation functions in the model have not been
studied analytically so far. In particular, the power spectrum of the
stochastic outflux of particles subjected to a steady slow input at top
layer seems to be an interesting quantity. There is fairly convincing
numerical evidence that this power spectrum is $1/f^2$ \cite{outflow}. It
would be useful to establish this theoretically. 

Another question related to time- dependent correlations is the
distribution of residence time of grains in the pile. This distribution
has been measured in experiments on ricepiles~\cite{residence}. The
directed model in two dimensions seems to be the simplest where the
properties of such distributions can be studied.

For the undirected BTW model in two dimensions, the most obvious unsolved
question is determination of avalanche exponents. 
We mentioned earlier that the result that the chemical distance exponent
for spanning trees in 2-dimensions is 5/4 comes from its connection to the
Potts model, and conformal field theory. The spanning trees problem is a
rather classical problem in graph theory. It seems desirable to have an
elementary (say combinatorial) proof of this result, which does not use
the heavy machinery of conformal field theory. One would also like to 
have similar results in higher dimensions.  More generally, most of the
textbook results about spanning trees deal with the number of trees of
specified local structure (some specified edges absent, or present).
However, questions about the large-scale geometrical structure of spanning
trees are
not so well studied \cite{2dtrees}.

\begin{figure}
\label{shape}
\begin{center}
\leavevmode
\psfig{figure=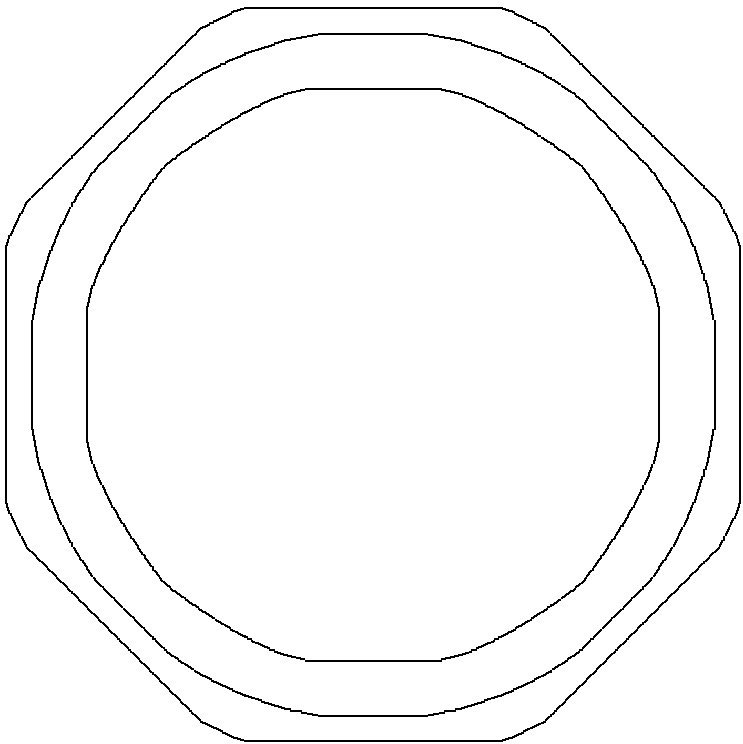,width=13cm,angle=0}
\end{center}
\caption{The shape of the base of the sandpile on a square lattice grown
by adding sand at one point. The three curve correspond to background with
all heights $0$ (innermost), a chessboard pattern formed with sites of
height $0$ and $1$, and all heights $1$ (outermost). The number of
particles added is $5 \times 10^5$ in each case.}
\end{figure}

Several other questions about the 2-dimensional model are rather
intriguing, and are not understood yet. One is the structure of the unique
recurrent state which corresponds to the identity element of the abelian
group. This shows interesting structures, at several length scales, and is
not understood. Some good pictures may be found in \cite{creutz}. Similar
structures are seen if the BTW model is relaxed from special unstable
states (like all heights $5$ \cite{gray}).  Similarly, we can consider
BTW model, with initial height zero everywhere, and add particles only at
the origin. As time evolves, the diameter of the region of space reached by the added
grains  increases as $t^{1/2}$. However, the asymptotic shape of
this region is not circular. If we grow the pile in a background of all
heights $1$, or on a chessboard pattern with half the sites of height $0$ and half of 
height $1$, the shape is also not circular, but a different shape. One clearly sees
some sharp facets [Fig. \ref{shape}]. An
understanding of how such shapes are selected in this simple model is
still lacking.

For the undirected ASM's in higher dimensions, not much is known. It is
usually believed that the upper critical dimension is $4$. This has
recently been shown by Priezzhev \cite{upperd}, using the equivalence to
the loop-erased random walk problem, and the fact that the upper critical
dimension is 4 for intersections of paths of random walkers. However, we
know that in the case of directed ASM, introduction of an arbitrarily
small stochasticity in the toppling rules can change the value of the
upper critical dimension from $3$ to $5$ \cite{tadic}. Whether a similar 
change in the upper critical dimension in the undirected case is not
settled yet.

There is some interest in understanding avalanche exponents on fractals
\cite{fractals}. 
The main reason for the interest is that one can hope to use exact
real-space renormalization group techniques. It turns out that abelian
property is
preserved under renormalization, but so far application of renormalization
group techniques to SOC has been somewhat ad-hoc \cite{RSRG}.
 
For $d \leq 4$, numerical simulations suggest that while avalanche clusters
can have nontrivial topology (two arms of propagating avalanche can join
to form a ring), the clusters are not very ramified, and the fractal
dimension of avalanche clusters is same as space dimension, i.e. number of
distinct sites toppled scales as $R^d$, where $R$ is the linear size of
the avalanche. It would be nice to have a proof (or disproof) of this
statement.

For the directed ASM, we showed that there is a simple systematic way to
exactly compute the probabilities of avalanches of sizes $1, 2 \ldots $.
This is not so for undirected models. The fact that one can calculate the
probabilities of different heights in the steady state implies that we can
calculate ${\sf Prob}(s=0)$. The calculation of ${\sf Prob}(s=1)$ already
involves the condition that all the neighbors of the toppled site are
strictly below the critical height, which has not been calculated so
far.

For the abelian models with stochastic toppling rules, very little is
known yet. The fact that introduction of stochasticity does not destroy
the abelian property seems to be important. It seems that some of the
treatment of the deterministic ASM's can be extended to the stochastic
case, and one can study the FSC's for the Manna model \cite{ddcalcutta99}.
However a variation of the burning test to determine whether a given
configuration is recurrent or not remains to be found for the Manna model.
Also, it is not established that the critical exponent of the 
Manna model are different from the deterministic case. In general, the
question of universality classes of the sandpile models is still open.

The relation of the avalanche propagation to propagation of infection in
contact process with many absorbing states is also quite unclear yet
\cite{dickman}. In general, the fact that avalanches correspond to
propagation of activity in a medium, where the number of absorbing states
of the medium is an exponentially increasing function of the number of
sites in it, is quite unexceptional. In the case of the stochastic
directed models studied by Tadic and Dhar \cite{tadic}, it was found that
the critical exponents with stochasticity are related to those of directed
percolation, but are `renormalized' by the conservation of mass condition.
Are the exponents of the undirected version of this model also related to
directed percolation exponents? An explicit solution would be very
illuminating , even if only in the 1-dimensional case.

I would like to thank Dr. Omri Gat, who prepared the first version of the
draft notes. I thank Professors Jean-Pierre Eckmann and Michel Droz of the
University of Geneva, and Professor Frederik Scholtz of the University of
Stellenbosch and the South African Organization of Theoretical Physicists
for inviting me to give these lectures, and Jean-Pierre Eckmann, Omri Gat
and Satya Majumdar for carefully reading the draft manuscript.


\begin{thebibliography}{999}
\bibitem{footnote2} In fact, we may assume that at each toppling, the
assignment of which grain goes to which neighbor is done randomly, then
the path of each grain from the point of addition to the boundary is a
random walk. In this case, the above bound can be improved to
$\left<s\right> \sim L^2$, which is actually an optimal bound.
\end{thebibliography}
\end{document}